\documentclass[review]{elsarticle}

\usepackage{lineno}
\usepackage[draft]{hyperref}
\modulolinenumbers[5]

\usepackage[T1]{fontenc}
\usepackage[utf8]{inputenc}
\usepackage{amsmath}
\usepackage{natbib}
\usepackage{graphicx}
\usepackage{booktabs}
\usepackage{gensymb}
\usepackage{url}
\usepackage{mathtools}
\usepackage{textcomp}
\usepackage[super]{nth}
\usepackage{chemformula}
\usepackage{multirow}
\usepackage[margin=2.5cm]{geometry}
\usepackage{esvect}
\usepackage{tabularx}
\usepackage{graphicx}
\usepackage{adjustbox}
\usepackage{siunitx}
\biboptions{authoryear}
\newcommand{\changemarker}[1]{%
\textcolor{black}{#1}%
}
\DeclareUnicodeCharacter{2212}{-}
\begin{document}

\begin{frontmatter}

\title{A Comparison of Fireball Luminous Efficiency Models using Acoustic Records}

\author[uwopa]{Luke McFadden}
\ead{lmcfadd6@uwo.ca}
\author[uwopa]{Peter G. Brown}
\author[uwopa]{Denis Vida}

\address[uwopa]{Department of Physics and Astronomy, University of Western Ontario, London, Ontario, N6A 3K7, Canada}

\begin{abstract}

The total energy of a fireball is commonly obtained from optical measurements with an assumed value for luminous efficiency. Acoustic energy measurements offer an independent means of energy estimation. Here we combine optical and acoustic methods to validate the luminous efficiency model of \citet{Borovicka2020}. Our goal is to compare these models with acoustic measurements of meteoroid energy deposition. Employing theoretical blast scaling laws following the approach of \citet{Mcfadden2021}, we determine explosive yields for both fireball fragmentation events and cylindrical shocks for four different bright fireballs. We model fireballs using the \texttt{MetSim} software \citep{vida2023direct} and find that the \citet{Borovicka2020} model produces agreement better than a factor of two for our three chondritic fireball case studies. The major exception is an iron meteorite-producing fireball where the luminous efficiency is an order of magnitude higher than model predictions calibrated with stony fireballs. We suggest that large disparities between optical and acoustic energies could be a signature of iron fireballs and hence useful as a discriminant of that population.

\end{abstract}

\end{frontmatter}


\section{Introduction}
\subsection{Motivation}

Meteoroids penetrate the Earth's atmosphere at velocities between Mach 30 and 240 \citep{Boyd1998, Ceplecha1998}, carrying substantial kinetic energy. This energy deposition can be on the order of kilotons of trinitrotoluene equivalent (1 kT TNT = $4.184 \times 10^{12} \si{\joule}$) for multi-meter sized objects. The majority of this energy is deposited in the atmosphere  \citep{Artemieva2016}, which is then converted to light, heat and sound. The most spectacular recent example of such a large fireball is the 2013 Chelyabinsk fireball. It inflicted significant damage to infrastructure and caused injuries to nearly 1500 individuals \citep{Brown2013} from the associated blast wave.

Impact risk assessment models crucial for planetary defense and decision-making heavily rely on fireball energy deposition measurements \citep{Reinhardt2016}. Numerous technologies are available to observe fireballs and estimate their energy deposition, including optical, radar, infrasound and other techniques \citep{Ceplecha1998}. For all observational techniques, however, a major limitation is determining the partitioning of total energy to that particular modality \citep{ReVelle2005}.  

Optical measurements are by far the most common method of observing fireballs. Ground-based optical networks currently cover a few percent of the Earth's atmospheric area, with camera systems such as the Global Meteor Network (GMN) \citep{Vida2021}, Cameras for All-Sky Surveillance (CAMS) \citep{Jenniskens2011a}, and the Global Fireball Observatory \citep{Devillepoix2020} in operation. Outside of dedicated networks, the improvement in consumer-grade camera quality and sensitivity in mobile devices, security cameras, and dashboard cameras have increased the amount of usable casual recordings of fireballs \citep{vida2021novo}. \par

Fireball optical intensity, spectroscopy data, and trajectory information can all be measured from video and camera imagery. These measurements usually consist of fireball position, speed, and brightness (light curve) versus either time, height, or path length since the beginning of the fireball. Fragmentation details can also be included if high-resolution video is available \citep{mcmullan2023winchcombe}. Based on these observations, a semi-empirical fragmentation model can be constructed reproducing the dynamic and photometric behaviour of the fireball, such as the \cite{Borovicka2013} model which has been used to reproduce observations of many meteorite-dropping fireballs \citep{Borovicka2020}. This model considers fragmentation as a discrete event which produces fragments that ablate independently. Critically, the model includes modes of fragmentation including dust: sudden dust release and continuous release of mm-sized particles from the fireball's surface (the process known as erosion) which has been found to be one of the most important factors in reconstructing fireball light curves \citep{shrbeny2020fireball}. The information from such modelling offers insight into the material properties of bodies impacting Earth, their strength and fragmentation behaviour all of which inform impact hazard models \citep{Mathias2017}. In this work, we use the \cite{vida2023direct} implementation of the \cite{Borovicka2013} model. \par

The optical energy produced by a meteor is a variable fraction of its total energy deposition. This fraction is called the luminous efficiency \citep{Ceplecha1998} and is often denoted as $\tau$. Empirical models \citep[e.g.]{Borovicka2020} have been derived that estimate $\tau$. Additionally, detailed theoretical models have recently become available for estimating $\tau$ with high precision for a limited set of speeds/masses \citep{Johnston2020}\changemarker{, as well as computational models with detailed simulations of hypersonic flow (see \cite{Dias2020} for more information and \cite{Popova2019} for a general review)}. The objective of this study is to verify the empirical luminous efficiency model independently through acoustic measurements.

Luminous efficiency relations can be sampled when modelling meteor ablation and fragmentation to produce a synthetic fireball light curve. An inverse procedure is applied to compute meteoroid initial mass from observations of the light curve and velocity \citep{vida2020new}.

Simply put, $\tau$ represents the ratio of emitted light power across all wavelengths and the total power of the meteor:
\begin{equation}
\label{eq:tau}
    \tau = \frac{I}{dE/dt}.
\end{equation}
Here, $I$ is meteor luminosity, and $dE/dt$ is instantaneous kinetic energy deposition per unit time \citep{Ceplecha1998}. Traditionally, the kinetic energy is used as a proxy for the total energy, as that is the only type of energy source a meteoroid is assumed to have. Luminous efficiency is a function of the meteoroid mass loss, deceleration, height, mass, velocity, and composition \citep{ReVelle2001}. Note that following \cite{ReVelle1980}, we refer to the instantaneous value for $\tau$ as the differential luminous efficiency, as opposed to the global trajectory-averaged value which we refer to as the integral luminous efficiency. In the following, where we do not specifically mention this distinction, we are referring to differential luminous efficiency.

The empirically-derived models which estimate instantaneous luminous efficiency take some of these variables into account, usually only the mass and velocity \citep{Ceplecha1976, ReVelle2001, Borovicka2013, Borovicka2020, vida2024first}. There have also been recent advancements in empirically validating the luminous efficiency based on composition, such as by \cite{Loehle2024}. However, due to the limited calibration datasets, owing to the rarity of well-observed large fireballs, the models have yet to be validated through independent approaches. \par

It is also possible to measure a time profile of instantaneous kinetic energy deposition for a fireball if acoustic signals are detected at the ground \citep{Edwards2009a}. This approach is an ideal way to validate existing fireball luminous efficiency models, provided simultaneous acoustic and optical records exist. 

The acoustic signal from a fireball may have two origins. As the fireball descends in the atmosphere, it produces a cylindrical shock which propagates outward and decays to lower amplitude pressure waves, the period and intensity of which are correlated with the energy deposition per unit path length along the fireball trajectory \citep{ReVelle1976a,Revelle1976b}. This outward propagating cylindrical shock is termed a ballistic wave. All fireballs produce ballistic shocks, though not all such signals reach the ground \citep{Edwards2009a}.

For most fireballs, one or more episodes of fragmentation occur during flight. Fragmentation results in sudden mass loss and a rapid increase in energy deposition. The resulting shock has a quasi-spherical geometry.

In practice, infrasound stations may record acoustic signals from both fragmentation episodes and a single arrival from the ballistic/cylindrical shock as described in \cite{Mcfadden2021}. This produces discrete measurements of energy deposition for multiple heights along the trajectory from a single station which can be directly compared to model predictions of luminous efficiency providing corresponding optical records exist, which is the goal of this paper. 


Adopting this approach, in this work we compare fireball energy deposition estimates from acoustic records with those constructed from optical estimates using the luminous efficiency model of \citet{Borovicka2020}. We incorporate this model of luminous efficiency into our software implementation of the semi-empirical fireball fragmentation model called \texttt{MetSim}\footnote{https://github.com/wmpg/WesternMeteorPyLib} which then produces a composite estimate for luminous efficiency weighted by the dust/fragment mass for all released material. Weighing by mass is important as the luminous efficiency of micrometer dust in the \citep{Borovicka2020}  $\tau$ model is half that of dm-sized bodies. This model source energy can then be directly compared to our measured acoustic energy estimates.

Here we examine four fireballs as case studies. We validate the empirically established luminous efficiency relationships proposed by \cite{Borovicka2020} by directly comparing them with source energies derived from shock wave theory \citep{KinneyGraham1985} for fragmentation events and cylindrical line source theory \citep{ReVelle1976a, Revelle1976b} for ballistic acoustic arrivals.

\section{Methodology}

\subsection{Event Selection and Data Sources}

Our focus is on larger fireballs where both photometry and near-field acoustic observations are available. We compiled a catalog of fireballs possessing documented photometry/energy data occurring in regions with dense infrasonic networks. Our event selection used sources including the Meteoritical Bulletin\footnote{https://www.lpi.usra.edu/meteor/}  (emphasizing fireballs associated with meteorite falls), data from US Government (USG) sensors published by the Center for Near Earth Object Studies (CNEOS)\footnote{https://cneos.jpl.nasa.gov/fireballs/}, the Geostationary Lightning Mapper (GLM)\footnote{https://neo-bolide.ndc.nasa.gov}, and NASA's Skyfall database\footnote{https://fireballs.ndc.nasa.gov/skyfall}.

We then cross-referenced the geographical coordinates of each fireball with infrasonic stations from the Incorporated Research Institutions for Seismology (IRIS)\footnote{https://ds.iris.edu/gmap/}, the Observations \& Research Facilities for European Seismology (ORFEUS)\footnote{https://www.orfeus-eu.org/}, and the infrasound component of the Comprehensive Nuclear-Test-Ban Treaty Organization (CTBTO)\footnote{https://www.ctbto.org/}. To be retained for analysis, a fireball had to have at least one infrasound detector situated within an approximate tolerance of 2 degrees in both geometric latitude and longitude from the reference coordinates of the meteor, as this is approximately the largest distance at which direct arrivals can be expected (about 150km away).

We refined the selection of approximately 20 fireball events since the year 2000 according to the following criteria:
\begin{enumerate}
    \item \textbf{Trajectory Accuracy}: The trajectory had to be reasonably well established by independent (usually ground-based) measurements.
    \item \textbf{Independent Light Curve}: Independent calibrated light curve data was required. This was either from GLM, USG or in some instances ground-based cameras. 
    \item \textbf{Clear Infrasound Arrival}: The infrasound data should clearly display an arrival signal associated with the meteor. Association was made by requiring the signal arrival to be consistent with direct arrival celerities of 0.28 - 0.33 km/s and arrival backazimuths within 5 degrees of the known fireball location. 
\end{enumerate}

Criterion 1 necessitated knowledge of either the start and end coordinates or at least one point and heading angles of the trajectory found independently by ground-based cameras. The precision of the trajectory did not have to be high, given that locating the acoustic source using ray tracing was constrained by imperfect knowledge of the atmosphere using the method detailed in Section \ref{sensitivity_measurement}.

Criterion 2 was applied since our work aims to offer an autonomous estimation of luminous efficiency when acoustic energy is known. Validation of luminous efficiency model estimates necessitates the availability of both optical and acoustic (total) energy. 

Criterion 3 was chosen due to the significant dependence of wave period and amplitudes on the acoustic yield and consequently the luminous efficiency. As such, we focused on data with high signal-to-noise ratios. It is also crucial to identify the segment of the meteor trail responsible for the acoustic signal, requiring a direct (and not ducted) arrival. This confines the usable range to typically under 150~km from the station \citep{Mcfadden2021}, therefore restricting the number of usable events further.

\subsubsection{CNEOS Data Description}

The USG sensors report data of bright fireballs occurring in Earth's atmosphere. This data includes a light curve, approximate geographical coordinates (latitude and longitude), elevation and timing of peak brightness. Additionally, an integrated energy value for the light curve and an estimation of the total impact energy (following the method outlined by \cite{Brown1996, Brown2002sat}) are provided. Two of our events have only USG lightcurves to constrain the optical light production. Some fireballs also have trajectory information in the form of x, y, z velocity components in a geocentric Earth-fixed reference frame.

Our analysis involves comparing the acoustic energy extracted from our infrasound analysis with this integrated luminous energy associated with the fragmentation in the light curve or (for ballistic arrivals) the energy deposited per unit path length at the specular point \citep[e.g.][]{Mcfadden2021}.

The dynamic data sourced from USG sensors is known to be of lower accuracy than ground-based measurements. As such, care must be exercised when using it for a comprehensive fireball analysis \citep{Devillepoix2019, Brown2023}. However, as we show later, these initial USG estimates suffice for our survey, as elaborated in Section \ref{sensitivity_measurement}.

\subsubsection{Ground-based Acoustic data}

We make use of infrasound data recorded both by infrasound arrays operated as part of the International Monitoring System of the Comprehensive Test-Ban Treaty Organization (CTBTO) \citep{Pichon2019} and individual microphones co-located with seismometers. 

The N4 network, under the management of the Albuquerque Seismological Laboratory (ASL), comprises of 145 currently active stations equipped with both seismometers and infrasound sensors. Situated in the Eastern United States of America, this dense network permits numerous independent acoustic detections of bolides occurring within this region. The N4 array is well covered by the ground footprint of the Geostationary Lightning Mapper (GLM) which identifies bolides while monitoring lightning activities across North and South America as well as the Pacific Ocean \citep{Jenniskens2018}. In addition, the popularity of the citizen science initiative, \textit{Raspberry Pi Shake \& Boom\footnote{https://raspberryshake.org/}}, gives supplementary seismic and infrasound stations typically in residential areas.

\subsection{Entry Modelling}

\begin{figure}
  \includegraphics[width=\linewidth]{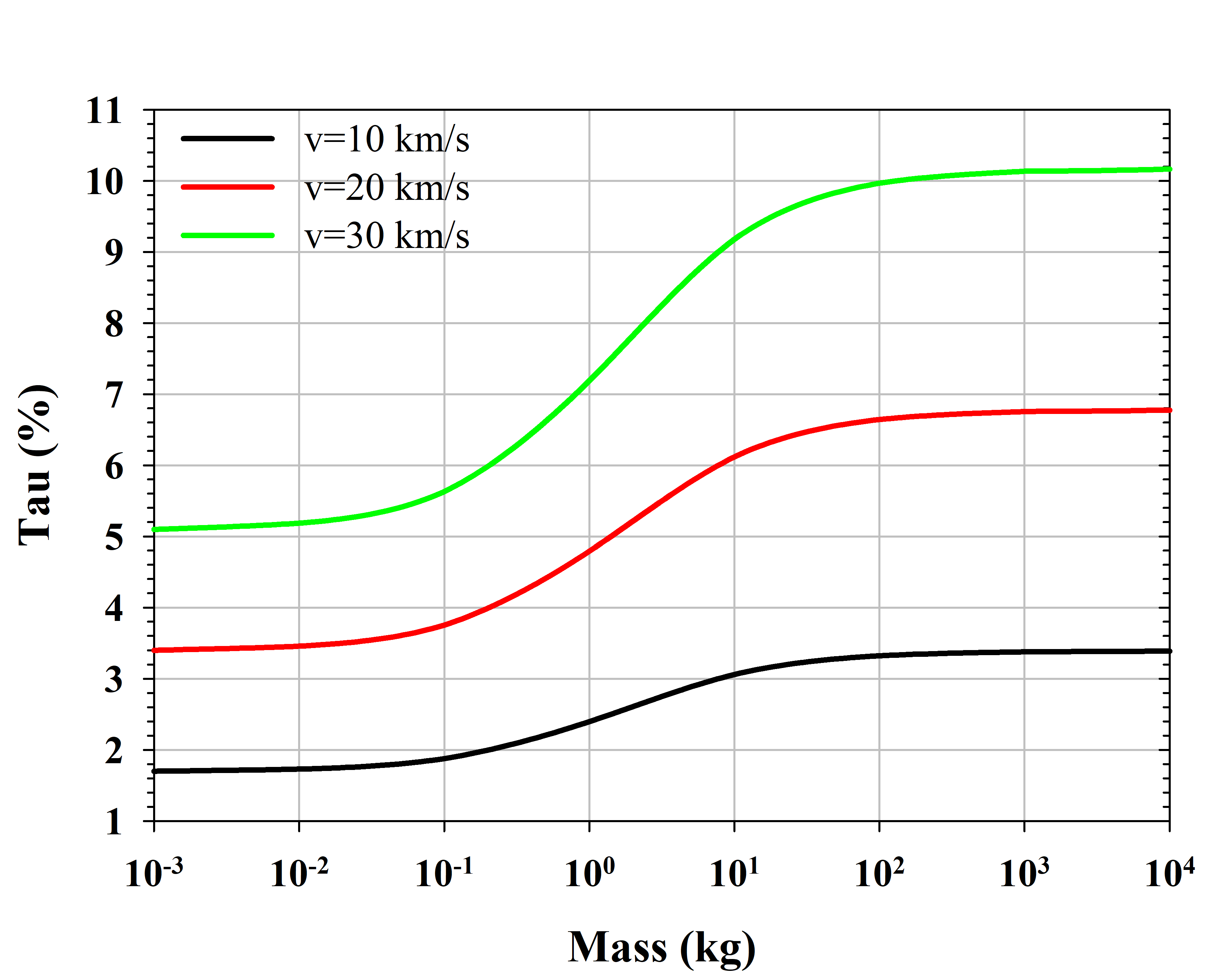}
  \caption{The mass and velocity dependence in the \cite{Borovicka2020} luminous efficiency model. Shown are the ranges expected for low speeds (<30~km/s) pertinent to this study.}
  \label{fig:borotau}
\end{figure}

Using the measured optical light curve and available dynamic information, we apply the \cite{vida2023direct} implementation of the semi-empirical fragmentation model of \cite{Borovicka2020} to estimate the energy deposition of each fireball. This model uses an empirical formulation of luminous efficiency with mass and velocity dependence as shown in Figure \ref{fig:borotau}. 

\cite{Borovicka2020} based their functional form for luminous efficiency on the velocity dependence proposed by \cite{Pecina1983} and the mass-dependence determined by \cite{ReVelle2001} based in part on the analysis presented in \cite{Ceplecha1996a}. Note that these original calibrations were validated based on the luminous efficiencies found for the Innisfree \citep{Halliday1981} and Lost City \citep{Ceplecha1996b} meteorite-producing fireballs, the only two meteorite-dropping fireballs where dynamics and luminosity were available at the time. The final version of the \citet{Borovicka2020} $\tau$ model also makes use of more recent meteorite-producing fireballs, (such as Ko\v{s}ice), to refine the final coefficients in the fit \citep{Popova2019}. The coefficients adopted in the model are appropriate to type I fireballs, i.e. stronger, chondritic objects which are expected to produce most observed meteorite falls \citep{Borovicka2020}.

The functional dependence shown in Figure \ref{fig:borotau} can be expressed as:

\begin{equation}\label{taulow}
\begin{split}
    \ln \tau = & 0.567 - 10.307 \ln V + 9.781(\ln V)^2 \\
    & - 3.0414(\ln V)^3 + 0.3213 (\ln V)^4 \\
    & + 0.347 \tanh(0.38 \ln M)\,,
\end{split}
\end{equation}

for velocities $V < 25.372$~km/s and:

\begin{equation}\label{tauhigh}
\begin{split}
    \ln \tau = -1.4286 + \ln V  + 0.347 \tanh(0.38 \ln M)\,,
    \end{split}
\end{equation}

for $V> 25.372$~km/s. Here M is the fragment mass in kilograms.

Within the \texttt{MetSim} software, a global average mass-weighted differential luminous efficiency is computed at each time step in the simulation, taking all fragments which produce light into consideration. This global $\tau$ thus takes into account the fragment mass distribution and the variation of $\tau$ with fragment mass and speed according to Eqs. \ref{taulow} and \ref{tauhigh}. This is then compared with our acoustically determined source energy, which is proportionate to the total energy deposition of the fireball. The model output from \texttt{MetSim} forms the template for our comparison with acoustic energy. It provides a model estimate for energy deposition and timing which we use for acoustic analysis. 

We manually adjust various fireball parameters to achieve an optimal fit to fireball observations (light curve and dynamics), keeping the observed initial velocity and entry angle \citep{vida2024first}. The adjusted parameters include the ablation coefficient, fragment mass loss, and whether fragments ablate as single bodies or through erosion and dust release. \changemarker{The changes in the ablation of both the main body and all modelled fragments are captured at distinct heights in the model. For simplicity and as there is limited observational information about the events, meteoroid physical properties are based on a rough classification of composition. More information about the ablation model used by \texttt{MetSim} can be found in \cite{vida2023direct}.} 

This manual fitting procedure is iteratively continued until the model light curve agrees with the observed light curve and dynamics (when available). Note that for the USG events no deceleration information is available and the resulting detailed fits are not unique, though the energy deposition (which is scaled through the luminous efficiency) closely matches the observations. \par 
\changemarker{The degeneracy of the fit is moderated by a few factors. First, we keep our fit parameters within physical limits found by \cite{Borovicka2020} who base their classification on hundreds of accurately measured fireballs. We note that the fit parameters are most ambiguous for the portions of the lightcurve where little to no fragmentation occurs. However, in the region near flares the characteristic fragmentation mode (eroding fragment, dust release, mass distribution of grains etc.) is relatively well defined by the height, amplitude, and duration of the flare, as noted by \cite{Borovicka2020}. We use these flares, where a substantial fraction of the total mass is often lost, to constrain our overall fits and reduce the degeneracy of the broad, non-fragmenting sections.}

As the first stage of modelling, the initial mass is first computed from the photometry using a fixed value of $\tau = 0.7\%$ as a starting value. The initial bulk density is chosen to either be chondritic ($\approx$ 3400~kgm$^{-3}$) or taken from the source references. The initial ablation coefficient is taken to be 0.005~kg/MJ \citep[as appropriate for chondritic fireballs ][]{Borovicka2013}, but may be adjusted during the modelling procedure. The entry angle is kept fixed at the observed value, after correcting it for the curvature of the Earth \citep{vida2024first}.

The nine variables that are manually estimated during the modelling procedure are:
\begin{enumerate}
    \item initial velocity $v_0$,
    \item initial meteoroid mass $m_0$,
    \item the height of each fragmentation $h_f$,
    \item the percentage of mass lost at the fragmentation point,
    \item the number of released fragments, $N$,
    \item the ablation coefficient,
    \item whether the fragments released are single bodies, lose mass through erosion, or are dust,
    \item for eroding fragments a grain mass distribution index between 1.0 and 3.0 is specified, with grain masses between $m_l$ and $m_u$,
    \item for eroding fragments the erosion coefficient is specified in kg/MJ.
\end{enumerate}

The final numerical parameters found to reproduce the lightcurve and available dynamics for each fireball case study can be found in \ref{MetSim_models}.

\subsection{Acoustic Energy}
\label{sec:acenergy}
Computing the acoustic source energy involves measuring the amplitude and dominant period of the received infrasonic signal. The methodology for energy computation hinges on whether the acoustic signal stems from the fireball ballistic (cylindrical shock) or fragmentation (spherical shock) source, which can be hard to distinguish in practice purely from the observed signal \citep{Edwards2009}. For a ballistic shock, the energy per unit path length is calculated following \cite{Silber2014}, while for fragmentation episodes, the total yield is determined as per \cite{Mcfadden2021}. This determination assumes knowledge of the trajectory, which for this study is mainly obtained from optical data, supplemented by acoustic records.

The source heights for each acoustic arrival are established by comparing the arrival times of the most prominent pressure signals in the infrasound waveform with ray-traced timings. These timings are derived using the known source time and position along the fireball path from optical records. For fragmentation events, seismic stations with corresponding signals are used in some cases to supplement the infrasound stations. In contrast, the height of the ballistic return received by a station is influenced by geometry, with distinct stations probing different heights of the fireball trajectory \citep{Mcfadden2021}.

Upon establishing the source heights through ray tracing and identifying the signal type (fragmentation or ballistic), the waveform underwent a three-step filtering process:

\begin{enumerate}
    \item \textbf{Removal of Mean}: To eliminate long-period amplitudes unrelated to fireball signals, the mean pressure amplitude was subtracted from the waveform.
    \item \textbf{Receiver Sensitivity Correction}: The waveform was adjusted for the responsivity of the receiver.
    \item \textbf{Bandpass Filtering}: The waveform was subjected to bandpass filtering within a range that yields the highest signal-to-noise ratio. Typically, this was found to be between 1 and 9 Hz.
\end{enumerate}

\begin{figure}
  \includegraphics[width=\linewidth]{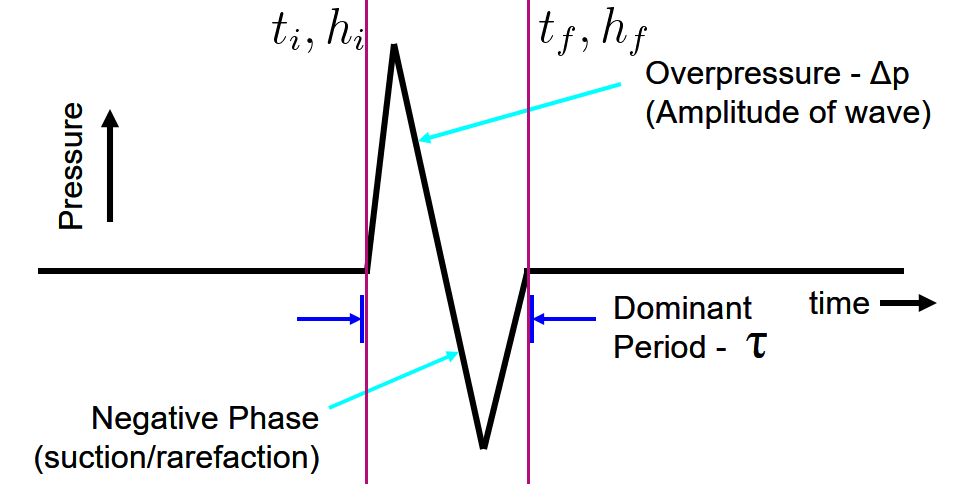}
  \caption{An idealized infrasound waveform generated from a fireball. $t_i$ and $t_f$ represent the beginning and end times of the waveform. Using the known fireball trajectory, it is possible to ray-trace from points along the trajectory until the travel times match with $t_i$ and $t_f$ to find their corresponding heights: $h_i$ and $h_f$.}
  \label{fig:wave_diagram}
\end{figure}

For each fireball-associated signal waveform, the beginning and end times of the signal were manually picked. As the signals we identify are all direct arrivals, we assume this temporal window can be taken as an upper limit to bound the fragmentation heights, recognizing that some of the spreading may be due to signal dispersion. This chosen range has the greatest impact on our results, as it directly defines the energy deposition integration range and subsequently the total energy. In addition, the period and overpressure at maximum amplitude are extracted for energy computations. Figure \ref{fig:wave_diagram} schematically illustrates the data measurement process for a representative waveform.

\subsection{Infrasound signal processing: Station Stacking}

For infrasound stations comprising of multiple elements (arrays), we sum the individual element waveforms based on the arrival's azimuth and altitude. This technique, waveform stacking, detailed in \citet{Ens2012}, enhances the signal-to-noise ratio. It presupposes a single source point for the signal and a planar arriving wave. In this stacking process, each station is temporally adjusted relative to a reference station, often the central element. The waveforms are then averaged using the linear stacking method found in the \texttt{ObsPy} package to mitigate noise \citep{Obspy}.

However, in fireball events featuring multiple acoustic sources, such as cases with several distinct fragmentations, the single source point assumption no longer holds. Therefore, the waveforms cannot be stacked simply, given that each fragmentation source necessitates a different time shift of order a few seconds, contingent on trajectory geometry. This was addressed by creating separate stacks for each source independently. Consequently, the resultant waveform displays favorable signal-to-noise characteristics exclusively for the chosen specific fragmentation.


\section{Determining fireball energetics from acoustic data}

\subsection{Meteor Acoustic Model}

\changemarker{In our acoustic model of a fireball, we assume that along the trajectory there are many points which release shock waves radially from the center, as a sphere. Since the meteor is travelling much faster than the speed of sound, adjacent points release shock spheres at approximately the same time. If the energy deposition is approximately the same between a section of points, then the resultant shape approximates a cylinder, due to superposition of multiple spherical shocks having the same blast radius between points. For most of the trajectory, the instantaneous energy deposition is not abruptly changing, allowing the cylinder assumption}. \par
\changemarker{In contrast, when fragmentation occurs, there is a sudden increase in energy deposition and the superposition of shocks having different blast radii emitted from adjacent points no longer cancel. This can be thought of as the geometry which results from a series of nested cylindrical blast radii all having different values across a small portion of the trajectory. The resultant shock shape is approximately a sphere (with minor interactions from nearby points) \citep{Silber2019}. This distinction is important because the geometry of observable locations for a spherical source is distinctly different than for a cylindrical source \citep{Mcfadden2021}}

\subsection{Spherical sources - fireball fragmentation energy: KG85 Method}
\label{fragenergy}

By measuring the infrasound amplitude at the ground originating from an atmospheric point-source explosion at an elevated position, it becomes possible to estimate the explosions' source energy via empirical fits to theoretical scaling relationships \citep{Hopkinson1915, Sachs1944, Kinney1968}. First, the scaled distance for an atmospheric explosion is defined as \citep{KinneyGraham1985}:

\begin{equation}\label{ScaledDistance}
    Z = f_d \frac{R}{{(W/W_0)}^{1/3}} \,,
\end{equation}

\noindent where $Z$ signifies the scaled distance (the distance away from an equivalent blast of a reference yield $W_0$), $f_d$ is a transmission term as defined in \cite{KinneyGraham1985}which takes into account the fragmentation height, $R$ represents the actual range between the source and the receiver, and  $W/W_0$ is the ratio of the actual yield $W$ to a reference yield $W_0$. \cite{KinneyGraham1985} provide equations linking scaled distance to overpressure $\Delta P$ and positive phase duration $t_d$, founded on experimental and modeling insights. This represents the resultant overpressure and phase duration after interaction with the atmosphere, including atmospheric dampening. Following \cite{Mcfadden2021}, we model fragmentations as chemical explosions. We use the empirical relation from \cite{KinneyGraham1985} to estimate overpressure as
\begin{equation}\label{chempressure}
    \frac{\Delta P}{P_R} = \frac{808 \left(1 + \left(\frac{Z}{4.5}\right)^2\right)}{\sqrt{1 + \left(\frac{Z}{0.048}\right)^2}\sqrt{1 + \left(\frac{Z}{0.32}\right)^2}\sqrt{1 + \left(\frac{Z}{1.35}\right)^2}} \,,
\end{equation}


\noindent where $P_R$ represents the ambient pressure at the reference station. From these equations, we may relate the overpressure observed at a station to the yield of an airborne explosion. \par

\subsubsection{A method for determining fragmentation energies in the absence of a full fireball trajectory }\label{sensitivity_measurement}

Large fireballs are rare and mostly occur in remote parts of the world. Thus they are poorly observed and their trajectories are not well known. As the blast equations rely on range estimates (such as Equation \ref{ScaledDistance}), this makes determining the fragmentation acoustic yield challenging. However, if two or more stations record a shared fragmentation event, the location of the fragmentation can be roughly constrained \citep{Edwards2004}. In practice, such localization approximations may still have uncertainties of order several kilometers. Our approach in such cases is to establish yield bounds given this location uncertainty.

Without precise knowledge of the fragmentation position, crucial variables such as the local ambient pressure at the burst height and the range – both integral in determining yield – are absent. To evaluate the change in yield within the uncertainties in range and height, we execute a straightforward sensitivity test.

This approach involves creating a three-dimensional grid encompassing latitude, longitude, and heights to simulate test fragmentation points, covering the full range of uncertainty. For each of these test points, we compute temporal error by ray-tracing the acoustic path to stations and then comparing the observed arrival time with the ray-traced time. We also compute the yield for these fragmentations from the observed overpressure. Subsequently, we create a three-dimensional map of latitude, longitude, and height, using the temporal error to all stations as an approximate metric of the likelihood of fragmentation occurrence at a particular point.

Considering a given temporal error tolerance (e.g., 3 seconds, representing a spatial error of about 1 km to account for sound speed), locations having timings errors less than this threshold are selected. The yield is then estimated for these points, with uncertainties derived using statistical distributions on a case-by-case basis. This process gives an expected yield value along with uncertainty bounds.

The procedure requires recording both temporal error and acoustic yield for each point, integrating attenuation as outlined by \cite{Mcfadden2021}. By allowing the temporal error a set tolerance and discarding points with temporal errors exceeding this threshold, we are left with points that define a two-dimensional area (latitude and longitude) of uncertainty in location. The energy values associated with each of these points reflect the extent of energy measurement spread. We find that the energy estimates adhere to a roughly normal distribution and the mean is considered our energy measurement, with the uncertainty defined as the range away from this mean, after removing outliers (refer to Section \ref{AlaskaCaseStudy} for more details).

Ideally, spatial data will converge towards a single point, and energy values will follow a normal distribution. However, this outcome may vary based on factors such as the number of stations utilized in simulations (more is better) and the particulars of the atmosphere. While these discrepancies were not evident in the presented case studies, they should be examined on a case-by-case basis. Moreover, this methodology integrates well with optical data. Cameras often provide the height of fragmentations as the brightest peak, permitting latitude and longitude search while maintaining a fixed height.

\subsection{Cylindrical line source fireball energy: Weak Shock Method}

For ballistic-source arrivals at our infrasound stations, we employ the Bolide Acoustic Modelling software package (BAM) \citep{Mcfadden2021}\footnote{https://github.com/wmpg/Supracenter}. We propagate acoustic rays radially outward from the fireball trajectory, with the initial launch angles perpendicular to the velocity vector (within a $25^{\circ}$ tolerance \citep{Brown2007}), to the ground. This creates an acoustic ground footprint from the trajectory, outlining a butterfly-shaped area (e.g. see Figure \ref{fig:romania_contour}) on the ground where it is possible to observe ballistic-source arrivals. This area is also called the boom corridor. Stations within the boom corridor may observe both fragmentation-induced spherical and ballistic-source arrivals, while stations outside it only observe fragmentations.  \par
\changemarker{Acoustic eigenrays are calculated with the Tau-P method following \cite{Edwards2003}. For the empirical weak-shock analytic formalism used in \texttt{BAM} as provided by \citep{ReVelle1976a}, only the travel times and ranges are required to estimate the ground footprint of the boom corridor and its associated amplitude. More complex computational fluid dynamics approaches \citep{Anderson2018, Nemec2021} include non-linear and destructive interference effects as they travel from source to receiver but at very steep computational costs. For BAM, the computational speed of the Tau-P method works well in concert with the analytic weak-shock formalism of \citep{Revelle1976b} allowing millions of ray-tracings along a trajectory to produce the ground footprint of a ballistic source. The ray-traces in BAM may also be perturbed by uncertainties in the atmospheric model to generate an estimated uncertainty in each value \citep{Mcfadden2021}.  Since we model a fragmentation (ballistic source) in BAM as a single point, the single binary arrival obtained by the Tau-P method is sufficient here. We note that the Cart3D approach of \cite{Anderson2018} was directly compared to the analytic weak shock results for several fireballs in \cite{Gi2018}, who found the ground overpressures differed by less than a factor of two. In future, we hope a CFD approach to shock propagation will be incorporated into BAM.}

Once the ballistic arrivals are identified, we calculate the dominant period using the zero-crossings of the waveform after it has been bandpassed to maximize signal-to-noise (generally about 1-9 Hz). This dominant period is input along with the source height into the \texttt{Geminus} program \citep{geminus}, which returns the relaxation radius, $R_0$, following the algorithm summarized in \cite{Edwards2009a}. Physically, $R_0$ can be viewed as a measure of the radial distance the shockwave moves before it is roughly slowed to the local speed of sound (See Figure 1 of \cite{Sakurai1964}). It is used to find the energy deposition per unit path length, as discussed below in Section \ref{tau_calc_section}.

\subsection{Luminous Efficiency Calculation} \label{tau_calc_section}

We can rewrite the differential luminous efficiency, as defined by \cite{Ceplecha1998}, from Eq. \ref{eq:tau} as:

\begin{equation}\label{eq:lum_eff_Intensity}
    I = \tau \frac{dE}{dt} \,,
\end{equation}

\noindent whereas before $I$ represents the luminosity, denoting optical energy per unit time, and $dE/dt$ stands for the total energy deposited by the fireball per unit time. The total energy deposited per unit path length for acoustically observed fireballs is estimated by computing the blast relaxation radius. For a cylindrical line source, the relaxation radius \citep{Edwards2008} is related to the energy deposited per unit path length $E_L$ through the equation:

\begin{equation}\label{eq:relaxation_traj}
    {E_L} = R_0^2 P \,.
\end{equation}

Here, $R_0$ is the relaxation radius and $P$ is the ambient pressure at the height of the source. Some references introduce a coefficient before this term, since this is a scaling metric for how quickly the shock wave speed reduces to the ambient speed of sound. For the full discussion, refer to \cite{Mcfadden2021, Silber2019}. For our purposes, we maintain the leading term as unity.

To estimate the differential luminous efficiency for a portion of a trajectory where a cylindrical shock is sampled by a ground station, denoted as $\tau_{ballistic}$, we combine Equations \ref{eq:lum_eff_Intensity} and \ref{eq:relaxation_traj}:

\begin{equation*}
    E_L = \frac{dE}{dl} = \frac{1}{v} \frac{dE}{dt} \,,
\end{equation*}


\begin{equation} \label{eq:traj_lum_eff}
    \tau_{ballistic} = \frac{I}{{dE}/{dt}} = \frac{I}{vR_0^2P} \,.
\end{equation}

This approach is suitable for the cylindrical or ballistic shock. For a fragmentation point, which we presume exhibits spherical symmetry, we define $\tau_{frag}$ and integrate Equation \ref{eq:lum_eff_Intensity} over the trajectory length of the fragmentation. For fragmentation starting at height $h_0$ and ending at height $h_1$:

\begin{equation*}
    \int^{h_1}_{h_0} I dh = \tau_{frag} \int^{h_1}_{h_0} \frac{dE}{dt} dh \,,
\end{equation*}
\begin{equation} \label{eq:frag_lum_eff}
    \tau_{frag} = \frac{\int^{h_1}_{h_0} I dh}{v_z E} \,.
\end{equation}

In the above, $\int^{h_1}_{h_0} I dh$ represents the area under a luminosity-height curve between $h_0$ and $h_1$, $v_z$ denotes the vertical component of the meteoroid's speed, and $E$ signifies the total energy released during the fragmentation. This notionally models fragmentation as a region along the trajectory where the relaxation (or blast) radius experiences an abrupt increase and fast decrease, approximating it as a spherical source. We may then use existing empirical relations between yields and ground-level pressure amplitudes of spherical explosions at altitude to gauge energy released as described in section \ref{fragenergy}.

Given that optical data is discrete in height measurements (e.g., specific shutter breaks or individual frames in a camera record), Equation \ref{eq:frag_lum_eff} can be simply approximated for discrete measurements using the trapezoidal rule as:

\begin{equation} \label{eq:frag_lum_eff_discrete}
    \tau_{frag} = \frac{1}{2v_zE}{\sum_{n = 1}^{N} (I_{n-1} + I_n)\Delta h_n} \,.
\end{equation}

Here, $N$ is the number of discrete observations along the light curve, $I_n$ is the luminosity at the height $h_n$, and $\Delta h_n$ is the difference in height between $h_n$ and $h_{n-1}$. This integration scheme is adopted in our software.

Applying Equations \ref{eq:traj_lum_eff} and \ref{eq:frag_lum_eff_discrete}, we can estimate the luminous efficiency of the fireball at the heights of the fragmentations and the source points along the trajectory. In theory, this permits generating a sample of $\tau_{ballistic}$ for each station recording a ballistic arrival. The source height of this arrival, derived from the trajectory, is contingent on good geometry between the fireball and the receiver. Additionally, for each acoustic arrival of a fragmentation at a station, an independent estimate is made for $\tau_{frag}$.

\subsection{Energy Conversions}

According to \citet{Ceplecha1998}, the relationship between luminosity $I$ and the observed absolute magnitude of the fireball $M$ is given by:

\begin{equation} \label{mag2lum}
    I = I_0 10^{-0.4 M} \,,
\end{equation}

\noindent where $I_0$ is a calibration zero point and is called the solid angle power of a zero-magnitude meteor (in W/ster). $I_0$ depends on the spectral bandpass of the sensor and the spectral energy distribution of the fireball. The $I_0$ which produces bolometric luminosity for fireballs observed by USG sensors is $I_0 = 248~\si{\watt}$/ster \citep{Brown1996} and Equation \ref{mag2lum} can be approximated as:

\begin{equation} \label{Brown1996eq}
    I = 10^{-0.4 (6 - M)} \,.
\end{equation}

Note that the reported USG sensor values on the CNEOS website assume a 6000~K black body spectral energy distribution \citep{Brown2002sat}. \changemarker{This 6000~K blackbody assumption is only a crude approximation to the true energy distribution \citep{Golub1997}, though it has been found empirically to match the total energy estimate found from infrasound estimates \citep{Gi2017}}. For other sensors, such as ground-based video cameras, appropriate values are used following \cite{WerykBrown2013}. The $I_0$ values used for each event are given in Table \ref{MetSim_init_params} as $I_0 = P_0/4\pi$.


\section{Results}

We analyzed four fireballs for which trajectory information, optical light curves and infrasound data were all available. Their details are given in Table \ref{tab:CNEOSdata} and include:

\begin{enumerate}
    \item A fireball which occurred over Romania on Jan 7, 2015 and had a ground track directly over an infrasound array. It had both CNEOS and ground-based light curve measurements \citep{Borovicka2017}.
    \item A fireball over Alaska on Feb 26, 2015 where coarse trajectory data from CNEOS was available together with a CNEOS light curve. The ground path was near a number of single infrasound microphones where the acoustic signal was clearly recorded.
    \item An iron meteorite-producing fireball over Sweden on Nov 7, 2020 detected by CNEOS \citep{Kyrylenko2023} and local ground-based cameras of the Norsk Meteor network producing acoustic signals recorded by a nearby Raspberry Boom station and several more distant infrasound arrays.
    \item A fireball over New York state on Aug 19, 2020 which was observed by several cameras of the Southern Ontario Meteor Network \citep{Brown2010a} and NASA All-sky fireball network \citep{Cooke2012}. This event was also detected at 8 nearby infrasound stations.
\end{enumerate}

\begin{table*}[!tbp]
\label{tab:CNEOSdata}
\begin{adjustbox}{width=\textwidth}
\begin{tabular}{lcccccl}

\hline\hline
Fireball & Peak Brightness [UTC] & Latitude {[}deg N{]} & Longitude {[}deg E{]} & Altitude {[}km{]} & Velocity {[}km/s{]} & \begin{tabular}[c]{@{}l@{}}Calculated Total \\ Impact Energy {[}kT{]}\end{tabular} \\
\hline
Romania  &  2015-01-07 01:05:59  &  45.7  &   26.9 & 45.5 & 35.7 & 0.40\\
Alaska   &  2015-02-26 22:06:24  &  68.0  & -149.0 & 33.7 & 21.1 & 0.53\\
Fröslunda   &  2020-11-07 21:27:04  &  59.8  &   16.8 & 22.3 & 16.7 & 0.33\\
\hline
\end{tabular}

\end{adjustbox}
\caption{A summary of the fireballs for which data from CNEOS is available. The calculated total impact energy assumes the relation between optical energy and total yield given in \citet{Brown2002sat} which corresponds to approximately an 8\% luminous efficiency for the entire trajectory. The New York State fireball was too faint to be detected by CNEOS sensors.}
\end{table*}

\subsection{Romania Fireball}\label{romanian_event}

\begin{figure}
  \includegraphics[width=\linewidth]{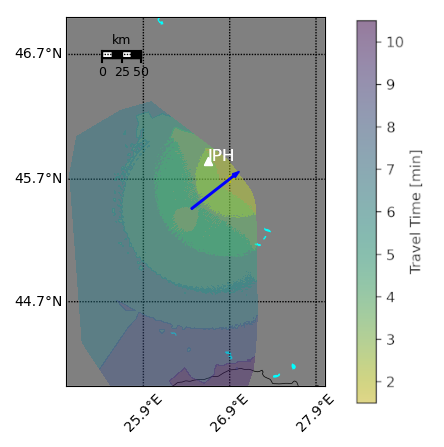}
  \caption{The ground footprint "boom" corridor for the Romania fireball (blue line), using the trajectory from \citet{Borovicka2017}. This shaded region shows the locations as a function of travel time (color coded) from the fireball where ballistic acoustic arrivals are expected to reach the ground together with the IPH infrasound array (red triangle). Since the RO-IPH array is within this boundary, it is likely that one of the infrasound arrivals detected at IPH is from the cylindrical shock, while other arrivals may be associated with one or more fragmentation episodes.}
  \label{fig:romania_contour}
\end{figure}

On January 7th, 2015, at 01:05:59 UTC, a superbolide was detected over Romania \citep{Borovicka2017}. It was captured by the European fireball network \citep{spurny2006automation}, US government sensors, and ground-based casual video recordings. The best estimate for the light curve for this event as reconstructed by \cite{Borovicka2017} is shown in the left inset of Figure \ref{fig:romania_lc_energy}. This fireball is of interest for our work as it is one of the few superbolides having a trajectory close to an infrasound array. It was only 25~km from the 6-element RO-IPH array, as illustrated in Figure \ref{fig:romania_contour}, which was well within the boom corridor.

\citet{Borovicka2017} analysed this fireball in detail from records of two all-sky cameras of the European fireball network and five casual video recordings from Romania. They compared these measurements to the velocity and light curve reported by CNEOS, finding a significantly lower speed than reported by CNEOS but similar total energy. They also found that the orbit was likely asteroidal, though near the boundary for objects with Jupiter-family comet parents. They also noted that the meteoroid disrupted severely at 1~MPa dynamic pressure and then completely disintegrated into dust and small fragments with no significant mass above a few grams surviving near 42~km height. They suggested the Romania fireball represents an asteroidal body that has a relatively weak global strength, with no components having compressive strengths above 1~MPa and is unrepresented in meteorite collections.

\begin{figure*}
  \includegraphics[width=\linewidth]{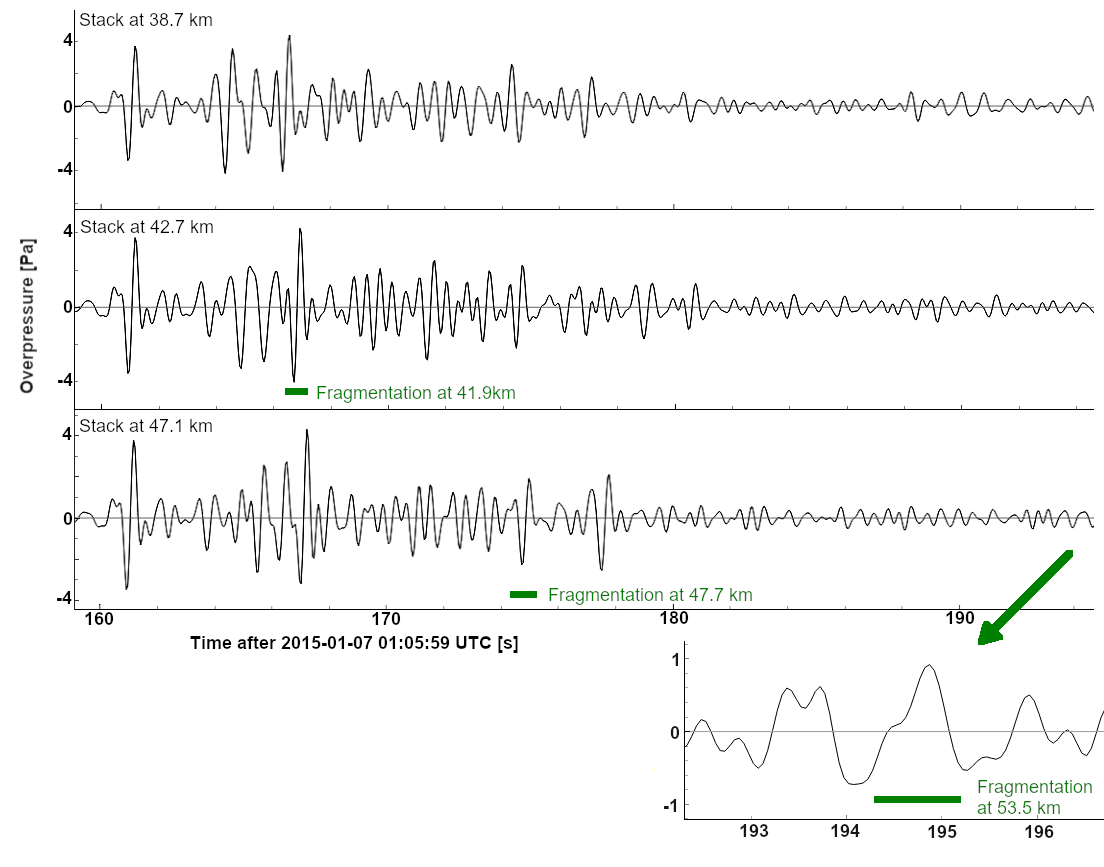}
  \caption{The pressure signal from the Romania fireball as a function of time as measured from the RO-IPH array, bandpassed from 0.5 - 3 Hz. Here each plot represents a different stack aligned at three different times to correspond to sources from heights of 38.7, 42.7, 47.1, and 53.5 km on the trajectory. Note the small difference in the amplitude of signals along the waveforms, depending on the stack. The timing of the fragmentations are labelled in green on the waveform they were measured from with the length of the green line indicating the length of the fragmentation.}
  \label{fig:romania_stack}
\end{figure*}

\begin{figure*}
  \includegraphics[width=\linewidth]{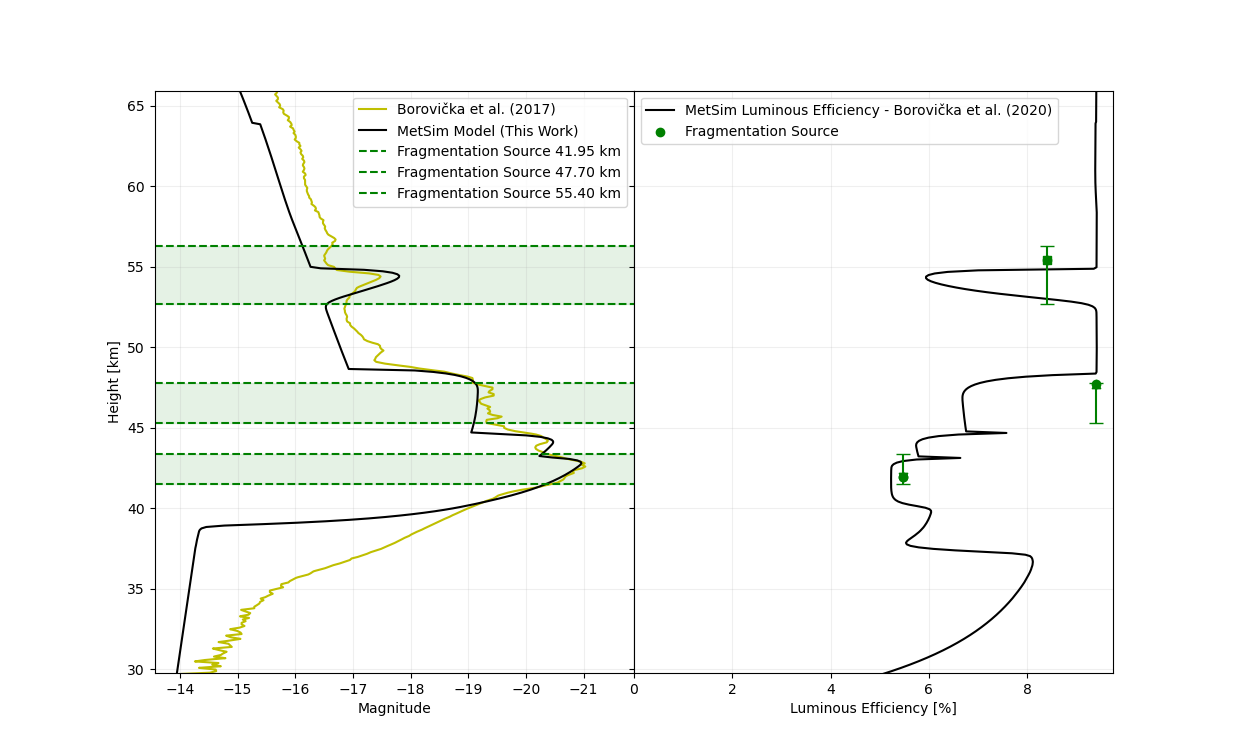}
  \caption{The light curve (left) measured by \citet{Borovicka2017} and the equivalent model mass weighted total luminous efficiency (right) as a function of height for the Romania fireball. The green dashed lines indicate the height boundaries of each fragmentation based on the duration of the individual acoustic signals. The points in the right plot represent the equivalent luminous efficiency from the acoustic records where the height range is represented by the error bars and the scatter point represents the height corresponding to the peak amplitude.}
  \label{fig:romania_lc_energy}
\end{figure*}

The infrasound record from the fireball is very complex at the IPH array. Examining each waveform individually, it became evident that there were four distinct acoustic sources at different heights, with the three fragmentations outlined in Table \ref{tab:Romania_Events}. Using the trajectory from \citet{Borovicka2017}, points along the fireball were ray-traced until the timing of the arrival matched the observed travel time. The waveforms for each stack are shown in Figure \ref{fig:romania_stack}, with stacks assuming sources along the trajectory at heights of 38.7, 42.7, and 47.1~km. Arrival times for the approximate beginning and end of signals within each waveform were recorded. Signals with common features across differing elements of the array were compiled and ray-traced to determine their arrival height.

The two initial arrivals from the lowest heights (38.4 and 41.9~km), had acoustic launch angle ($\phi$) relative to the perpendicular to the flight direction below $25^{\circ}$. This is the value determined by \citet{Brown2007} to be the limiting angular deviation which can still be consistent with a ballistic shock.  This suggests these returns could either be ballistic or from fragmentation. 

Conversely, the last two arrivals from the largest heights (47.7 and 55.4~km) were deemed to be from fragmentation episodes, as the deviations were well above the $25^{\circ}$ limit. A closer examination of the light curve showed no flares around 38.4~km, with the main one near 42.7~km and smaller ones at 47~km and 54~km, consistent with this interpretation. This indicates that the 38.4~km source is likely not associated with a fragmentation episode as it lacks an optical flare. Thus we interpret the 38.4~km arrival as ballistic, while the subsequent ones were identified as discrete fragmentations. We also associate a height interval with each return based on the duration of the wave train around the signal maximum. 

Based on fragmentation and ablation model parameters found by \cite{Borovicka2017}, we reproduce the light curve and dynamics in \texttt{MetSim} but using the \cite{Borovicka2020} luminous efficiency model. The left inset of Figure \ref{fig:romania_lc_energy} shows the comparison between the observed and simulated light curve, together with the acoustically observed heights of fragmentation. The right inset of the figure shows the model luminous efficiency weighted by the mass of fragments producing light (smaller fragments have a smaller luminous efficiency) compared to luminous efficiencies derived using the acoustic source energies at the four source heights within the observed height boundaries of each source, (see \cite{Vida2021} for more information).

Fragmentation returns are illustrated in Figure \ref{fig:romania_lc_energy} as areas beneath the light curve, with dashed lines indicating the equivalent height range based on the start and end timing of each arrival on the waveform stack. The first major fragmentation modelled by \citet{Borovicka2017} near 53.5~km corresponds to our highest acoustically observed fragmentation. The model $\tau$ is around 9.41\%, while our estimate is 8.4\%. The fragmentation arrival at 47.7~km is not directly observed in the light curve as a singular peak, but \cite{Borovicka2017} modelled the sudden rise in the light curve between 48 and 42~km as a series of a dozen rapidly occurring smaller fragmentation events. Our acoustically inferred luminous efficiency of 9.40\% is reasonably close to the model value of 6.95\%, while matching the model from the same height. The 41.9~km arrival matches well with the final fragmentation which produced a flare after which the fireball rapidly decreased in brightness. Our value of $\tau = 5.60\%$ matches well with the model value of 5.63\%.

If we assume that most of the meteoroid's energy is deposited in the fragmentations, we arrive at a total acoustic energy of 0.45~kT TNT equivalent, very similar to both the CNEOS estimation of 0.40~kT and the \cite{Borovicka2017} value of 0.41~kT. This agreement confirms the validity of the \cite{Borovicka2020} luminous efficiency model for this fireball.

\begin{table*}[!tbp] 

	\centering 
	
	\begin{tabular}{l c c c c c c l} 
	\hline\hline 
	Height [km] & $\phi$ [deg] & $\Delta P$ [Pa] & $T$ [s] & $t_d$ [ms] &  Yield [kT TNT] & $R_0$ [m] & Acoustic $\tau$ [\%]\\ 
	\hline 
	$41.9^{+1.5}_{-0.4}$ & 22.54 & 5.45  & 0.58 & 322.4 & 0.360 & 2000 & 5.60\\
	$47.7^{+0.1}_{-2.4}$ & 33.43 & 2.10  & 1.27 & 573.3 & 0.074 & 1500 & 9.40\\
	$55.4^{+0.9}_{-2.7}$ & 40.91 & 0.77  & 0.57 & 283.9 & 0.013 & 800 & 8.40\\
	\hline 
	\end{tabular}
	\caption{Observations of the three distinct signals in the RO-IPH array stack. The source height is calculated using the arrival time at the station relative to the fireball timing (from the stack with the closest source height to the ray-traced height). The superscript and subscript values indicate the bounds of the fragmentation. Here $\phi$ is the angle between the perpendicular to the velocity vector and the initial launch vector of the ray-trace solution, $\Delta P$ is the measured overpressure of the maximum amplitude at the array using a bandpass with the highest SNR (typically values of 0.5 - 3 Hz), $T$ is the pressure period at maximum amplitude, and $t_d$ is the duration of the first phase. $R_0$ is the relaxation radius for the ballistic arrival is calculated following \citet{ReVelle1976a} and for the fragmentation-related signals using the expression in \citet{Mcfadden2021}.}
    \label{tab:Romania_Events}
\end{table*}


	
	

\subsection{Alaska Fireball} \label{AlaskaCaseStudy}

On February 26th, 2015, at 22:06:24 UTC, a fireball occurred over Northern Alaska that was detected by USG sensors. Several seismic and one microbarometer station are located within a few hundred kilometers from the fireball. The measured fireball data from USG are summarized in Table \ref{tab:CNEOSdata}. However, as no ground-based trajectory records exist, we note that the $\approx$10~km precision given by CNEOS for the latitude and longitude of the fireball was insufficient for ray-tracing solutions. Consequently, the alternative method, described in Section \ref{sensitivity_measurement}, was employed to constrain USG-only trajectories.

\begin{figure}
  \includegraphics[width=\linewidth]{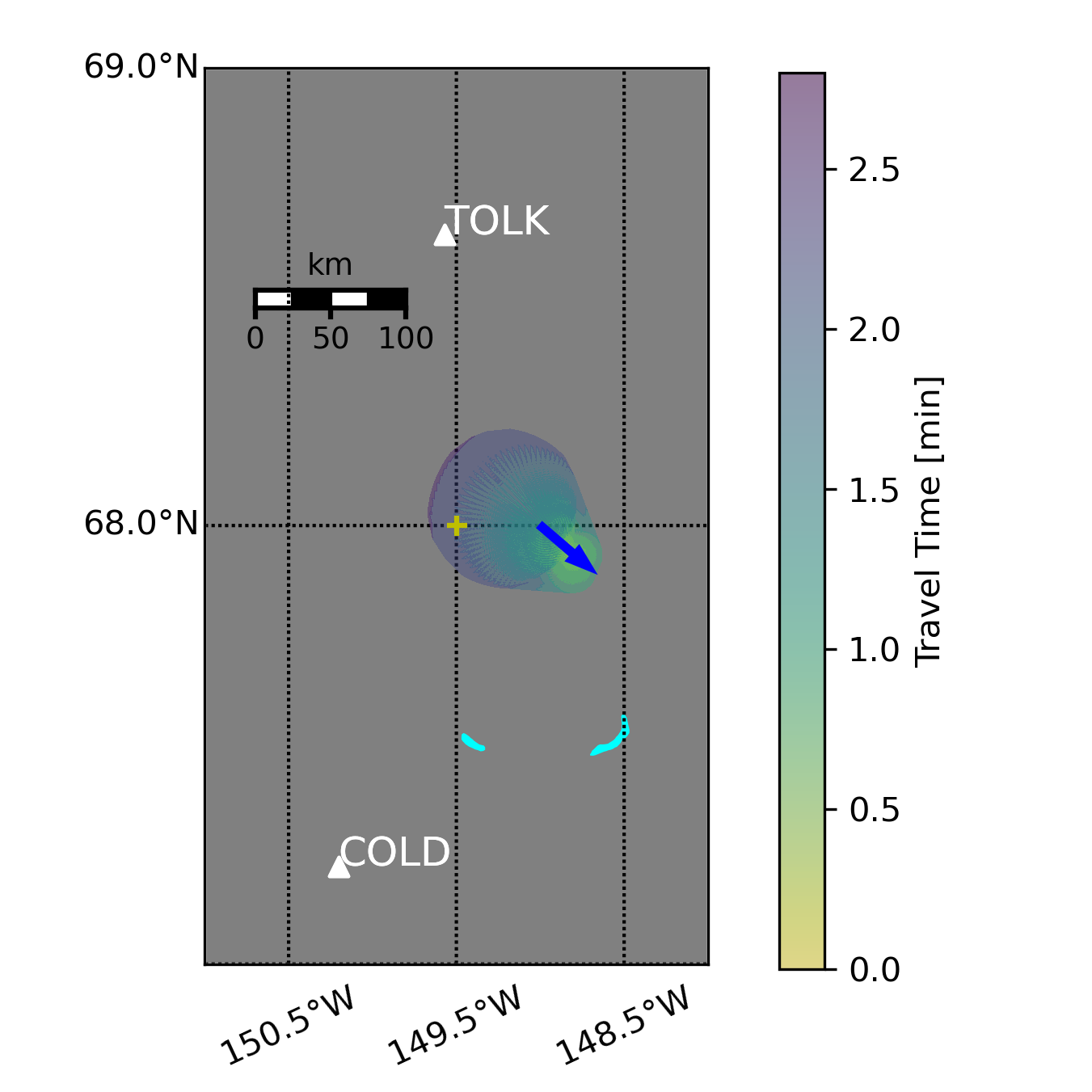}
  \caption{The boom corridor for the Alaska fireball. The ground trajectory (as given by CNEOS) is shown in blue, while the region where ballistic acoustic arrivals reach the ground is the shaded region. Perpendicular arrivals from the trajectory did not reach either TA-TOLK (seismic and infrasound station) or AK-COLD (seismic only) to be observed. We therefore infer that signals received were likely from fragmentations only. The origin of the local coordinate system used is shown as a yellow cross.}
  \label{fig:alaksa_contour}
\end{figure}

\begin{figure*}
  \includegraphics[width=\linewidth]{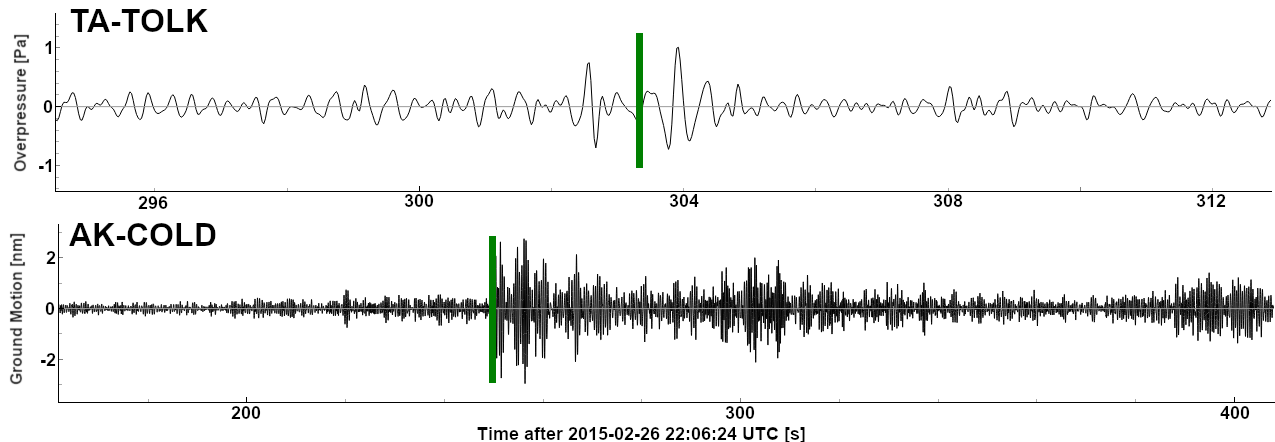}
  \caption{The time series for the two near-field stations of the Alaska event. TA-TOLK (top) was the only available near-field infrasound sensor, while the seismic station AK-COLD (bottom) was used to constrain the position of the fragmentation Supracenter. Both were bandpassed from 2-8 Hz to optimize for signal-to-noise. The green vertical line represents the time of the pick used.}
  \label{fig:alaska_waveform}
\end{figure*}

The boom corridor associated with this event is depicted in Figure \ref{fig:alaksa_contour}. Two stations were within a range of 150~km of the trajectory and both acoustically observed the fireball: TA-TOLK, which includes both a pressure sensor and a seismic sensor, and AK-COLD, a seismic station fortunately positioned on the opposite side of the trajectory. This allowed for a better geometrically defined fragmentation analysis. 

As ballistic arrivals were not expected to reach either station (as they were outside of the boom corridor) we presumed a single fragmentation created the observed signals. To better localize the fragmentation point, a two-dimensional grid search was conducted for the fragmentation location with a constant height of 33.7~km which is the height at peak brightness reported by CNEOS. The time difference between the observed and ray-trace transit times from each point to each station was used as a measure of the goodness of fit. The height at peak brightness reported by CNEOS has been found from other comparisons with ground-based records to be accurate to 2-3~km in most cases \citep{Brown2023}. This search was guided by timing data from both stations, pressure measurements from TA-TOLK and the known time of the fireball from CNEOS records.

A broad search area in the region between ($151.0^{\circ}$W, $147.0^{\circ}$W) and ($66.0^{\circ}$N, $69.0^{\circ}$N) at the fixed height was explored. For each point in the search, the temporal error for both stations and the yield for a fragmentation based on the infrasound record at TA-TOLK were computed.

\begin{figure}
  \includegraphics[width=\linewidth]{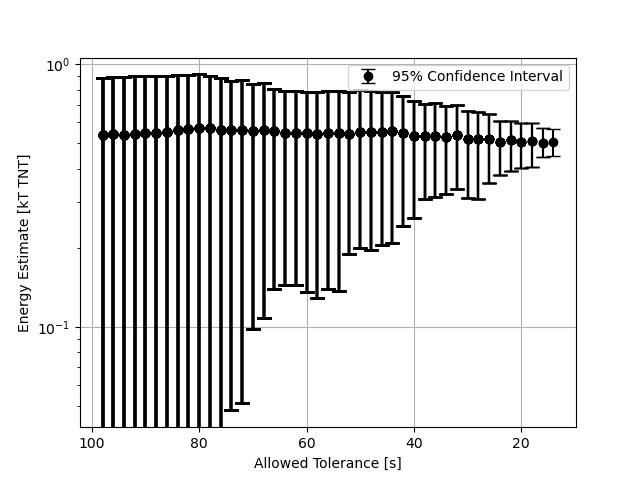}
  \caption{Since the uncertainty in yield in the ray-tracing solution may be smaller than the yield uncertainty caused by atmospheric uncertainties impacting ray-tracing estimates, a solution with a slightly higher error may be a better model of reality than a solution with a low ray-tracing error. This plot shows the range in energy estimates under a given ray-tracing error threshold. Since the energy converges as we reduce the tolerance, we can estimate a reasonable value for the energy deposition for this event. The mean solution with each step is shown as the point for each tolerance, and the error bars represent the outer limits of the 95\% confidence interval of the data available with a tolerance less than the given step. These percentiles were chosen instead of the range so that outlier values from erroneous ray traces near the edges of the grid did not impact results, i.e. removing data with unphysically large or small energies.}
  \label{fig:alaksa_frag_error}
\end{figure}

\begin{figure}
  \includegraphics[width=\linewidth]{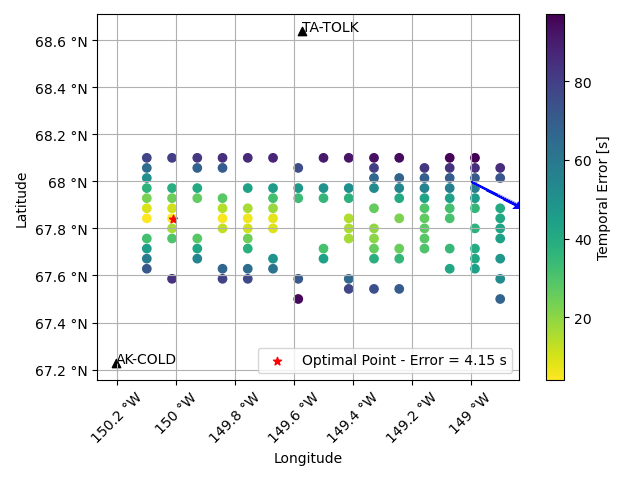}
  \caption{The timing uncertainty as a function of position for the Alaska fireball. This complements Figure \ref{fig:alaksa_frag_error}. Fragmentation sources are shown as scatter points, with their ray-tracing timing error shown by their colour. As the error tolerance decreases, the number of points also decreases, and ideally should converge to one location, which would be our spatial uncertainty for the fragmentation location or supracenter. The blue arrow represents the approximate USG trajectory.}
  \label{fig:alaksa_frag_final}
\end{figure}

\begin{figure*}
  \includegraphics[width=\linewidth]{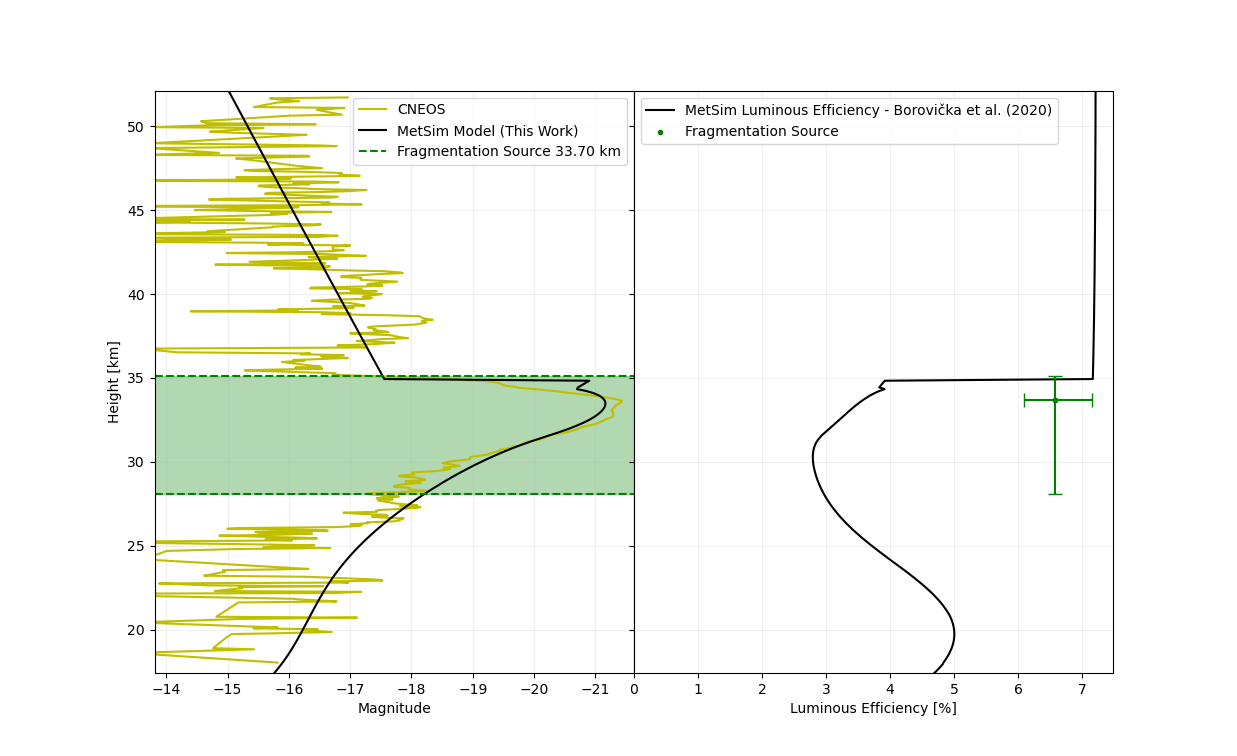}
  \caption{The light curve for the Alaska fireball measured by USG, here converted into height using the information from the CNEOS website for velocity and altitude of the brightest peak (left). The dashed green box indicates the height range of the fragmentation corresponding to the duration of the acoustic signal. The total luminous efficiency from all contributing fragments modelled using the \cite{Borovicka2020} model  is shown for comparison to the right.}
  \label{fig:alaksa_lc}
\end{figure*}

Considering solely the fragmentation location error, a yield of $0.51^{+0.05}_{-0.04}$~kT TNT (1 standard deviation from the mean) was found for fragmentation locations having a temporal error of 14 seconds or less (the temporal range with 10 grid search points remaining). Convergence of the energy estimate with lower tolerances is depicted in Figure \ref{fig:alaksa_frag_error}. Optimal positions for the fireball are displayed in Figure \ref{fig:alaksa_frag_final}, with the simulation's best point being $67.84^{\circ}$N, $150.01^{\circ}$W for the CNEOS reported height of 33.7~km, a temporal error of 4.15~seconds, and a yield of 0.51~kT TNT. This energy estimate is consistent with CNEOS's estimate of 0.53~kT TNT and with our interpretation that the main fragmentation deposited almost all the fireball energy, a conclusion also supported by the fireball light curve as shown in the left inset of Figure \ref{fig:alaksa_lc}.

As with the Romania fireball, we used the CNEOS speed and entry angle together with the observed light curve to model the fireball and its energy deposition. The fireball was modelled as a rocky body with an ablation coefficient of $\sigma = 0.005$kg/MJ and fragmentation at 34.5~km in height. Details about the modelling can be found in \ref{MetSim_models}.

From our source energy estimate of $0.51^{+0.05}_{-0.04}$ kT TNT  and the observed light curve, we find a luminous efficiency of $6.2 \pm 0.5$\%  at the point of the flare near 33~km. This luminous efficiency is compared to the total luminous efficiency generated by the fragmentation model, as illustrated in the right inset of Figure \ref{fig:alaksa_lc}.

If we allow for the fragmentation as having occurred in a height range of 30.0 - 36.0~km (with a grid spacing of 0.1~km), the optimal solution shifts to $67.84^{\circ}$N, $149.93^{\circ}$W, at a height of 30.3~km with a temporal error of 0.26 seconds. For this solution, we find a yield of 0.24~kT. However, while this solution has minimal time residuals with only two stations and given atmospheric propagation uncertainties the potential errors in ray-tracing alone are on the order of a few seconds \citep{Mcfadden2021}. Reproducing our grid-tolerance search using a tolerance of 10 seconds we find a range of source yields from $0.46^{+0.40}_{-0.24}$ kT TNT. This variation in yield corresponds to a possible range in luminous efficiency from 3.67\% to 14.3\%, with our best estimate remaining 6.5\%.

While we conclude that there is agreement within error between our acoustic energies and the model luminous efficiency, this case study illustrates the sensitivity in yield values which results when even modest errors of order a few kilometers are present in a fragmentation position. It also highlights the limitations imposed by the coarse positional information in USG measurements.

\subsection{Fröslunda iron meteorite-dropping fireball}

\begin{figure*}
  \includegraphics[width=\linewidth]{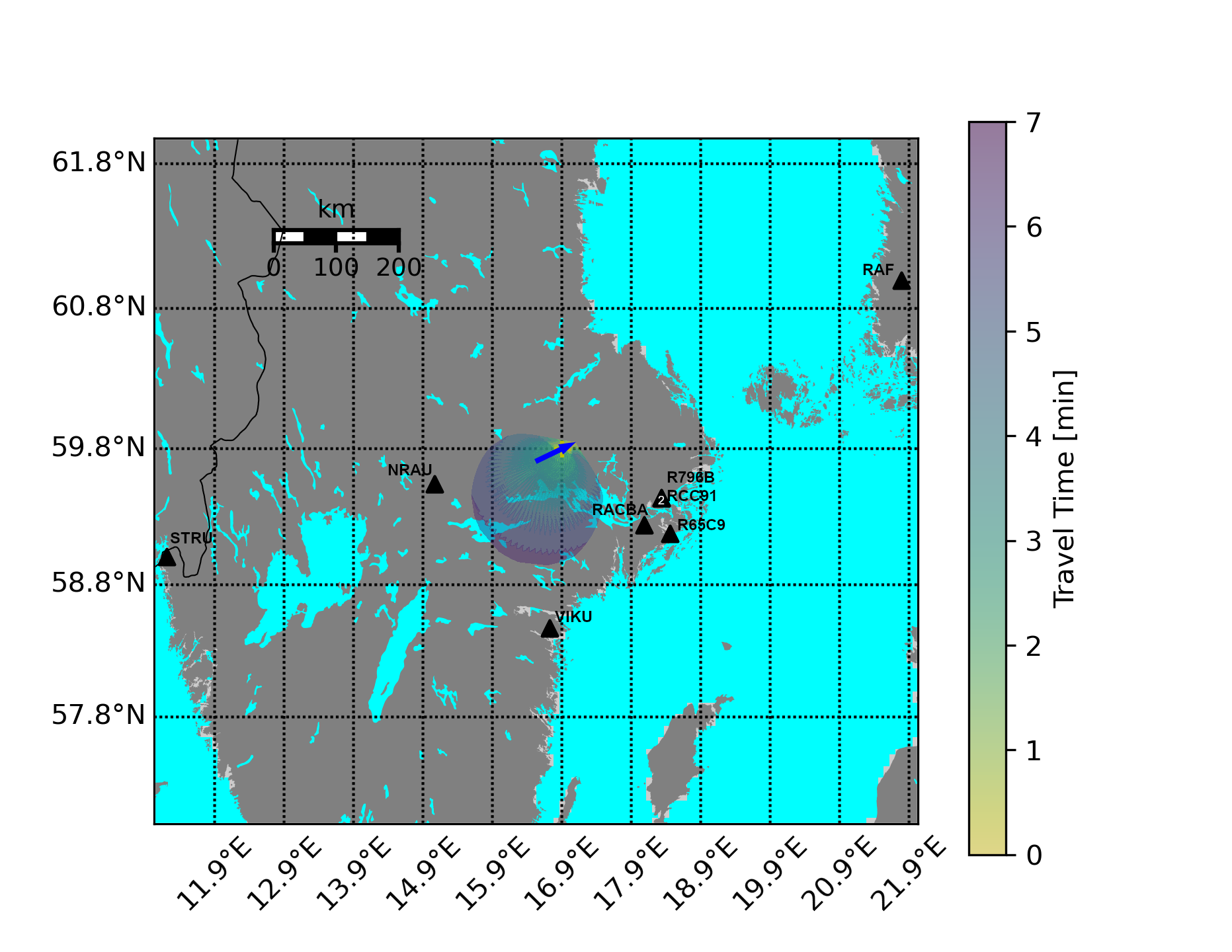}
  \caption{The "boom" corridor for the Fröslunda event. At the time of the event, strong upper winds pushed the acoustic rays toward the South East. Stations to the West, such as NRAU, are unable to observe any meteor-related acoustic signals. Since the wind was strong, it is possible that the "boom" corridor does not fully capture the extent of the ballistic rays in the downwind direction.}
  \label{fig:Sweden_Ground_Track}
\end{figure*}

The Fröslunda Fireball occurred on November 7th, 2020, at 21:27:04 UTC over Sweden and resulted in the fall of a 13.8~kg iron meteorite \citep{Kyrylenko2023}, as shown in Figure \ref{fig:Sweden_Ground_Track}. Fireball data are available from video records of the Norsk Meteor Nettverk\footnote{http://norskmeteornettverk.no/meteor/20201107/212700/} as well as from CNEOS. This fireball is of particular interest due to its iron composition, a relatively rare occurrence for bright fireballs as a result of which no empirical/observational constraints for fireball iron luminous efficiency are available.

A mass of approximately 3540~kg was inverted in the modelling, assuming a bulk density of 7000~kg/m$^3$, appropriate to iron. We also used a high ablation coefficient of $0.1$~kg/MJ consistent with the large mass loss rate experienced by irons as a result of melting \citep{Ceplecha1998} instead of vaporization experienced by rocky meteoroids. We note that due to the different modes of ablation and mass loss experienced by iron meteoroids, our fragmentation model is not fully appropriate. However, as we are only interested in reproducing the zeroth-order energy deposition of the fireball, we judge that our model is sufficient if it can match the light curve.

\begin{figure*}
  \includegraphics[width=\linewidth]{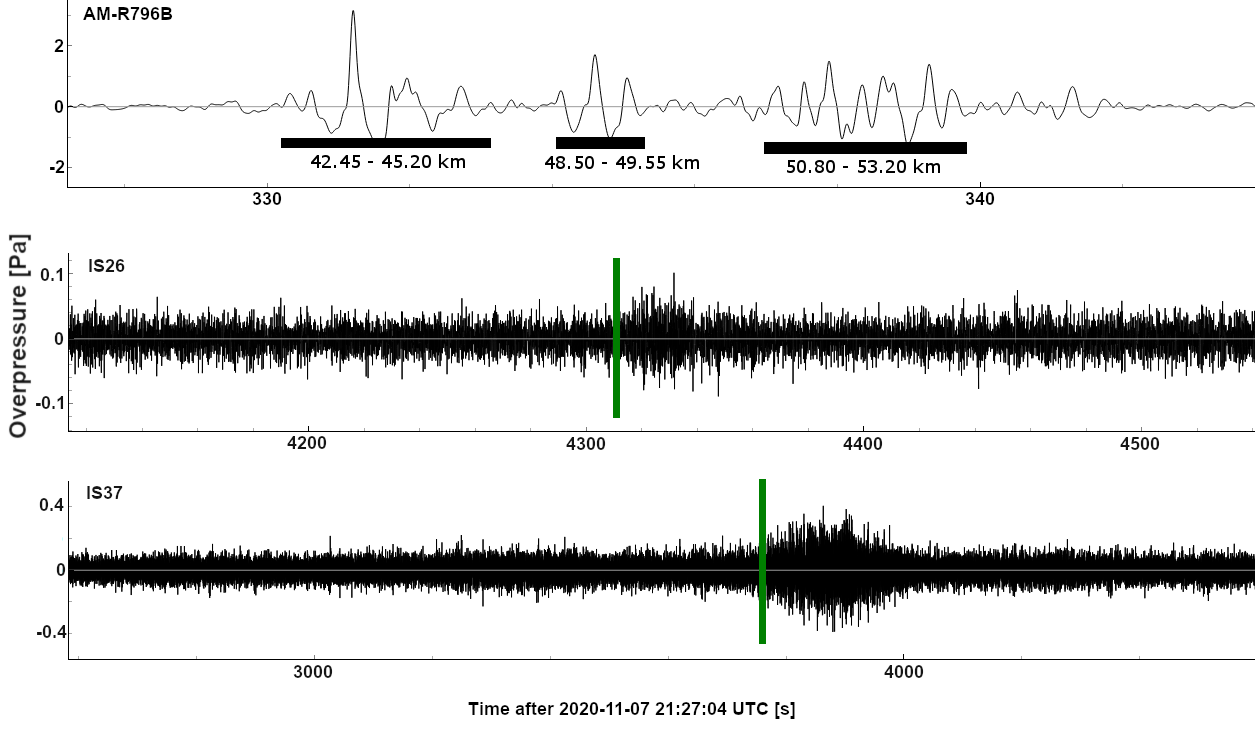}
  \caption{The acoustic signal as recorded at station AM-R796B less than 100~km range from the Fröslunda fireball. A 1 - 9 Hz bandpass Butterworth filter was applied to the raw waveform. The three signals indicated by the dark black lines are interpreted to be fragmentations with the labelling indicating equivalent heights along the trajectory. These heights were found by ray-tracing from the known trajectory and finding which heights corresponded with the observed arrival times. The pressure signals obtained from far-field CTBTO stations IS26 in Germany (middle) and IS37 in Norway (bottom), as used in Table \ref{estimated_energy_tab} for the Fröslunda event bandpassed from 0.5-4 Hz. The arrival occurs near 22:42 UT at IS26 and at 22:32 UT at IS37. The maximum amplitude and corresponding dominant period are extracted following the procedure described in \citep{Ens2012} and used to estimate total fireball energy and therefore luminous efficiencies as summarized in Table \ref{estimated_energy_tab}. The green vertical line represents the approximate signal start time for the far-field stations}
  \label{fig:norsk_infra}
\end{figure*}

The energy estimate was derived from near- and far-field infrasound stations (shown in Figure \ref{fig:norsk_infra}) and is summarized in Table \ref{norsk_infrasound_energies}. Note that the far-field stations provide a global (integral) luminous efficiency \citep{ReVelle1980} for the entire event, while near-field records can provide estimates for the differential luminous efficiency of discrete portions of the fireball path.

We ray-trace the acoustic signals to the station from each given height indicated by the arrival times. The launch angles of the later two signals are above the tolerance of $\approx 25^{\circ}$, and can therefore be considered fragmentations over a ballistic arrival. The first signal's launch angle was $\approx 16^{\circ}$, which allows it to be ballistic or a fragmentation. Here we have assumed it to be a ballistic-source arrival, with the "boom" corridor from Figure \ref{fig:Sweden_Ground_Track} unable to reach the station because of a strong wind to the southeast. 

The calculated total energy from the near-field (direct) acoustic amplitude measurements, presuming the signals represent fragmentation points at these stations, follows the methodology from section \ref{sec:acenergy} and is presented in Table \ref{norsk_fragmentations}. For comparison, CNEOS reported a total impact energy of 0.33 kT TNT assuming an integral luminous efficiency of 8\%. As all our acoustic source energy estimates are lower than the CNEOS energy value, our measurements thus suggest the actual integral luminous efficiency is much higher than the \citet{Borovicka2020} model values, presuming the optical energy computed under the usual assumptions (see \citet{Brown2002sat} for a summary) is correct. We discuss further modelling details in Section \ref{discussion_summary}.

\begin{figure}
  \includegraphics[width=\linewidth]{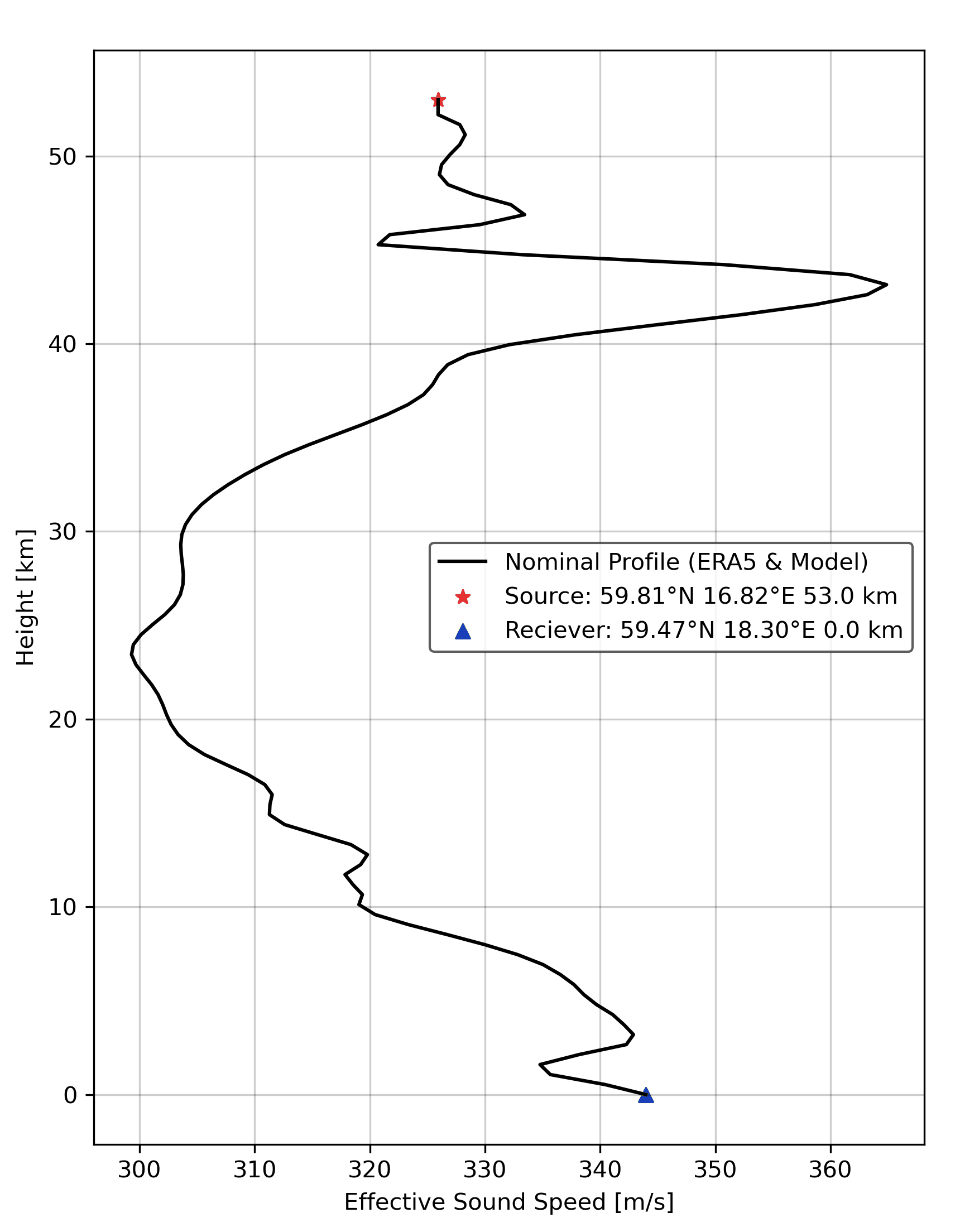}
  \caption{The effective sound speed profile from the Fröslunda fireball, based on the temperature and wind profiles from a source at 53.0~km altitude along the fireball trajectory to the station AM-R796B. The black line shows the weather data from ERA5 \citealp{Dee2011}, using data from NRLMSISE-00 \citep{NRLMSISE} and HWM \citep{HWM14} to extrapolate data above the 0.1~hPa level. The sound speed profile is such that sources below ~40~km height will produce acoustic rays that will not reach the ground. These reflected rays are not direct arrivals, and therefore are not useable for our near-field energy calculations. It also implies that any acoustic sources below 42 km are in a shadow zone as seen from AM-R796B.}
  \label{fig:norsk_atmos}
\end{figure}
\begin{figure*}
  \includegraphics[width=\linewidth]{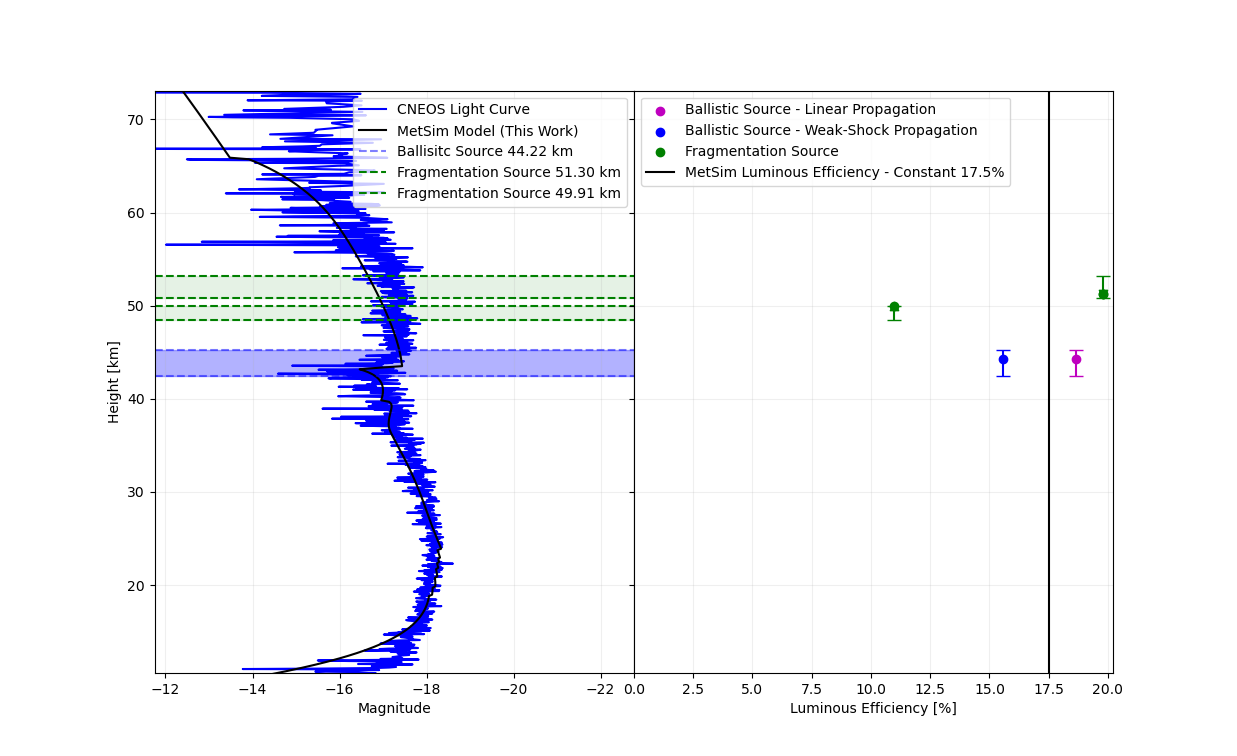}
  \caption{The light curve of the Fröslunda Fireball from CNEOS records (left) shown in blue together with the best-fit model lightcurve (black) using $\tau = 17.5\%$. Also shown are the height ranges corresponding to the acoustic sources, interpreted as fragmentations, based on arrival times from the near-field station AM-R796B (dashed lines). The equivalent luminous efficiencies produced from the ablation model and from the discrete The duration of the fragmentations is given by the boundaries of each fragmentation given in Figure \ref{fig:norsk_infra}.}
  \label{fig:norsk_lc_model}
\end{figure*}

The light curve for the fireball from CNEOS records is displayed in Figure \ref{fig:norsk_lc_model}. Observed acoustic arrival times from the one nearby infrasound station were confined to source regions around 50~km height; no signals were detected below 40~km. This may be due in part to the sound speed profile, as shown in Figure \ref{fig:norsk_atmos}. Since the speed of sound for acoustic rays near the ground is higher than it is for launch points \changemarker{along the trajectory} below 40~km, rays would tend to reflect horizontally, and potentially reflect upwards, by Snell's law. Figure \ref{fig:norsk_infra} shows the waveform captured by the only near-field infrasound sensor, as well as the two far-field detections. Three distinct wave trains were chosen and associated with fragmentations, as listed in Table \ref{norsk_fragmentations}. Notably, the light curve extends to lower altitudes than usually recorded for fireballs. It is estimated that the light curve is representative of the fireball to around 15~km altitude, below which the CNEOS observations are embedded in noise.


Inversion techniques from \cite{Edwards2003} to independently isolate fragmentation points were insufficient to accurately constrain the fireball due to the geographical distribution of seismic and infrasound observing stations. These stations were all positioned to the East of the trajectory and in close proximity. This results in an unconstrained supracenter solution due to the degeneracy between source time and 3D station distance.

Using the trajectory published by the Norsk Meteor Nettverk \citep{Kyrylenko2023} as a constraint, we searched for points along the trajectory which best explain the observed arrivals. Ray-tracing was employed from a 1D grid of heights along the trajectory to the AM-R796B infrasound station. The height interval corresponding to the start and end of each distinct signal at the infrasound station was determined and shown in Figure \ref{fig:norsk_infra}. The beginning and ending of each signal are used as a proxy for the range of heights covered by the fragmentation. Fragmentation energy for each fragmentation was calculated using the time of the largest amplitude of the signal. This information is presented in Figure \ref{fig:norsk_lc_model} and Table \ref{norsk_fragmentations}.

Our results suggest significantly higher luminous efficiencies for this iron fireball than predicted by the \citet{Borovicka2020} model, which was developed for typical chondritic meteoroids. For the integral luminous efficiency of the event as a whole, far-field stations produce a best-estimate of the yield from the two station average of the period of 1.5~s of 0.02~kT TNT, similar to the wind-corrected amplitude yield from the downwind I37NO station. The IS26 signal is much weaker, with the signal barely at the level of the noise as it is a counter-wind return and hence the amplitude yields are much less reliable. Under the assumption of a 6000~K black body, the CNEOS optical energy is roughly the same, indicating an unphysical value of $\tau$ near 100\%.

For the fragmentations, we find luminous efficiencies from the near-field observations of order 10-20\% as illustrated in Figure \ref{fig:norsk_lc_model}. Note that systematic shifts in the timing/height produce similar values for luminous efficiency as the light curve is quite flat between 45 - 55~km height. We also note that the acoustic signal associated with the fragmentation between 50.80 - 53.20~km, in particular, might produce an overestimate for $\tau$, as the observed acoustic signal in this range was quite lengthy and noisy, potentially resulting in an overestimate of the fragmentation duration. Using a fixed average $\tau$ of 17.5\%, we find an initial mass of the meteoroid to be around 3500~kg, and are able to reproduce the final meteorite mass to an order of magnitude, with our model value of 47.4~kg.


\begin{table*}[]
\label{norsk_infrasound_energies}
\begin{tabular}{@{}llllll@{}}
\toprule
Station  & \begin{tabular}[c]{@{}l@{}}Range \\ {[}km{]}\end{tabular} & \begin{tabular}[c]{@{}l@{}}Wind-corrected \\ Amplitude Yield\\ {[}kT TNT{]} \end{tabular} & \begin{tabular}[c]{@{}l@{}}Wind-corrected \\ P-P Amplitude Yield \\ {[}kT TNT{]}\end{tabular} & \begin{tabular}[c]{@{}l@{}} Period at \\ Maximum Amplitude \\ {[}s{]} \end{tabular} & \begin{tabular}[c]{@{}l@{}}Integral\\ Luminous\\ Efficiency {[}\%{]}\end{tabular} \\ \midrule
AM-R796B & 95                                                        & 0.24                                                                                     & 0.23  &  0.50                                                                                      & 11.1 - 11.5                                                                         \\
I37NO    & 1035                                                      & 0.02                                                                                     & 0.10  &  1.43                                                                                   & 26.5 - 133                                                                          \\
I26DE    & 1234                                                      & 0.004                                                                                    & 0.003 & 1.56                                                                                  & -                                                                                   \\ \bottomrule

\end{tabular} \label{estimated_energy_tab}
\caption{The total estimated energy of the Fröslunda fireball for both near- and far-field infrasound stations, determined using the wind-corrected amplitude of the observed signal and period found using the relations from \citet{Ens2012} (see their Table 2). The CTBTO stations were bandpassed from 0.6 - 8 Hz. The CNEOS yield is 0.33 kT TNT, which assumes an 8\% luminous efficiency along the whole trajectory. The equivalent luminous efficiencies from acoustic yields relative to the CNEOS lightcurve are shown in the final column.}
\end{table*}

\begin{table}[]
\begin{tabular}{@{}llll@{}}
\toprule
Height [km] & Source Type & Energy [T TNT]  & Tau [\%] \\ \midrule
42.45 - 45.20 & B & 0.09 /m &  15.6 - 18.6    \\
48.50 - 49.91 & F & 4.78 & 11.0  \\
50.80 - 53.20 & F & 4.42 & 19.8    \\ \bottomrule

\end{tabular}
\caption{Heights of ballistic (first line) and fragmentation (last two lines) identified from ray-trace timing for the Fröslunda fireball and their associated energies from station AM-R796B. The equivalent luminous efficiency using the CNEOS light curve is also shown. For the ballistic arrival, a line energy density is given.}
\label{norsk_fragmentations}
\end{table}
\subsection{New York State Fireball}

A modest $-8$ peak magnitude fireball occurred on August 19, 2020, in upstate New York, at 01:45:56 UTC. This event was observed by numerous all-sky camera systems, security cameras, and ground-based observers. We consider it in our work as 8 infrasound stations are located within a 200~km range of the trajectory. This potentially allows well-constrained uncertainties for energy and luminous efficiency measurements across a range of heights and from multiple independent acoustic records.

This fireball was not bright enough to be registered either by GLM or USG sensors. We reconstructed its light curve and trajectory from ground-based all-sky camera data of the SOMN and the NASA all-sky network \citep{Cooke2012}. The calculated trajectory, along with the boom corridor and location of nearby infrasound stations, are depicted in Figure \ref{fig:ny_map}. The trajectory solution from all three cameras is given in \ref{NY_fireball_traj_sol}. The measured light curve is presented in Figure \ref{fig:ny_lc_wmpl}, and enlarged in Figure \ref{fig:ny_lc_wmpl_zoom}. The light curve shows a very clear major fragmentation episode at 49~km altitude, after which the fireball rapidly dimmed. The light curves from cameras 4A and 17A matched well, but the light curve from 16A was brighter, having a systematic shift in brightness of 0.5 magnitudes due to poor observing conditions (low elevation, presence of anthropogenic light gradients).

\begin{figure*}
  \includegraphics[width=\linewidth]{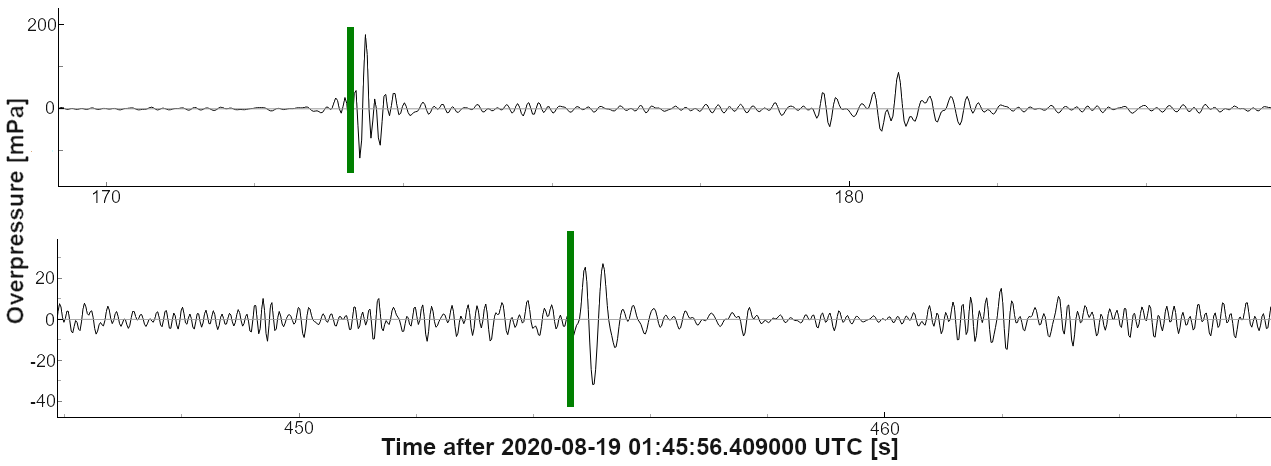}
  \caption{The infrasound signals associated with the New York state fireball. The top plot shows the signal from station N4-L56A while the bottom plot is the pressure-time series from N4-K57A. Both plots are bandpassed from 1-9 Hz. Two distinct arrivals are apparent at both stations as described in the text. The picks used are marked as green vertical lines.}
  \label{fig:NY_waveforms}
\end{figure*}

The data from two infrasound stations (Figure \ref{fig:NY_waveforms}) that recorded clear signals from the 49~km fragmentation are given in Table \ref{NY_infra_data}. We also verify that the launch angle was greater than $25^{\circ}$ away from perpendicular to the trajectory, likely due to strong winds, which eliminate the ballistic source approach. Both independently observed an almost identical yield of $\sim0.46$~kg TNT at different ranges (59 vs. 150~km) and recorded overpressures (100 vs. 40~mPa), confirming the accuracy of the empirical blast scaling laws for fragmentations proposed by \citet{Mcfadden2021}. This match has been facilitated by the similar launch angles of the acoustic rays from the fireball fragmentation source, making the waves experience comparable atmospheric effects.

\begin{table*}[htbp!]
\begin{tabular}{@{}lllll@{}}
\toprule
Station & \begin{tabular}[c]{@{}l@{}}Range to Fragmentation \\ Point @ 48.8 km height {[}km{]}\end{tabular} & \begin{tabular}[c]{@{}l@{}}Arrival Time \\ {[}s after reference{]}\end{tabular} & \begin{tabular}[c]{@{}l@{}}Overpressure \\ {[}mPa{]}\end{tabular}  & \begin{tabular}[c]{@{}l@{}}Yield \\ {[}kg TNT{]}\end{tabular} \\ \midrule
N4-L56A & 58.6 & 173.56 & 100 & 0.464 \\
N4-K57A & 150.2 & 454.73 & 40 & 0.459 \\ \bottomrule
\end{tabular}
\caption{Acoustic signal characteristics from the New York state fireball as recorded at two infrasound stations in the near-field regime showing the highest signal-to-noise ratio. The waveforms are bandpassed (typically between 1 - 9 Hz) to optimize for signal-to-noise to obtain the overpressure (amplitude). The reference time here is 01:45:56 UT on Aug 19, 2020.}
\label{NY_infra_data}
\end{table*}

\begin{table*}[htbp!]
\begin{tabular}{@{}lllllll@{}}
\toprule
Camera & \begin{tabular}[c]{@{}l@{}}Peak Height \\ {[}km{]}\end{tabular} & \begin{tabular}[c]{@{}l@{}}Peak Absolute \\ Magnitude {[}mag{]}\end{tabular} & \begin{tabular}[c]{@{}l@{}}Cut off \\ Magnitude {[}mag{]}\end{tabular} & \begin{tabular}[c]{@{}l@{}}Height Range \\ of Fragmentation {[}km{]}\end{tabular} & \begin{tabular}[c]{@{}l@{}}Optical\\ Energy {[}kJ{]}\end{tabular} & \begin{tabular}[c]{@{}l@{}}Luminous\\ Efficiency, $\tau$ {[}\%{]}\end{tabular} \\ \midrule
Cam004A & 48.82 & -7.98 & -7.38 & 48.64 - 49.05 & $99.8$ & 5.58 - 5.64\\
Cam017A & 49.00 & -7.96 & -7.37 & 48.59 - 49.14 & $126$ & 4.95 - 5.01 \\ \bottomrule
\end{tabular}
\caption{Luminosity data for the New York State fireball. Here the cut-off magnitude corresponds to the brightness when the flare was at 25\% of its peak luminosity. The corresponding height ranges are also shown. Note that the yield measurement in Table \ref{NY_infra_data} is multiplied by the instantaneous velocity in order to compare with the optical energy in the flare which in turn is used in finding luminous efficiency.}
\label{NY_lum_eff_data}
\end{table*}

\begin{figure*}
  \includegraphics[width=\linewidth]{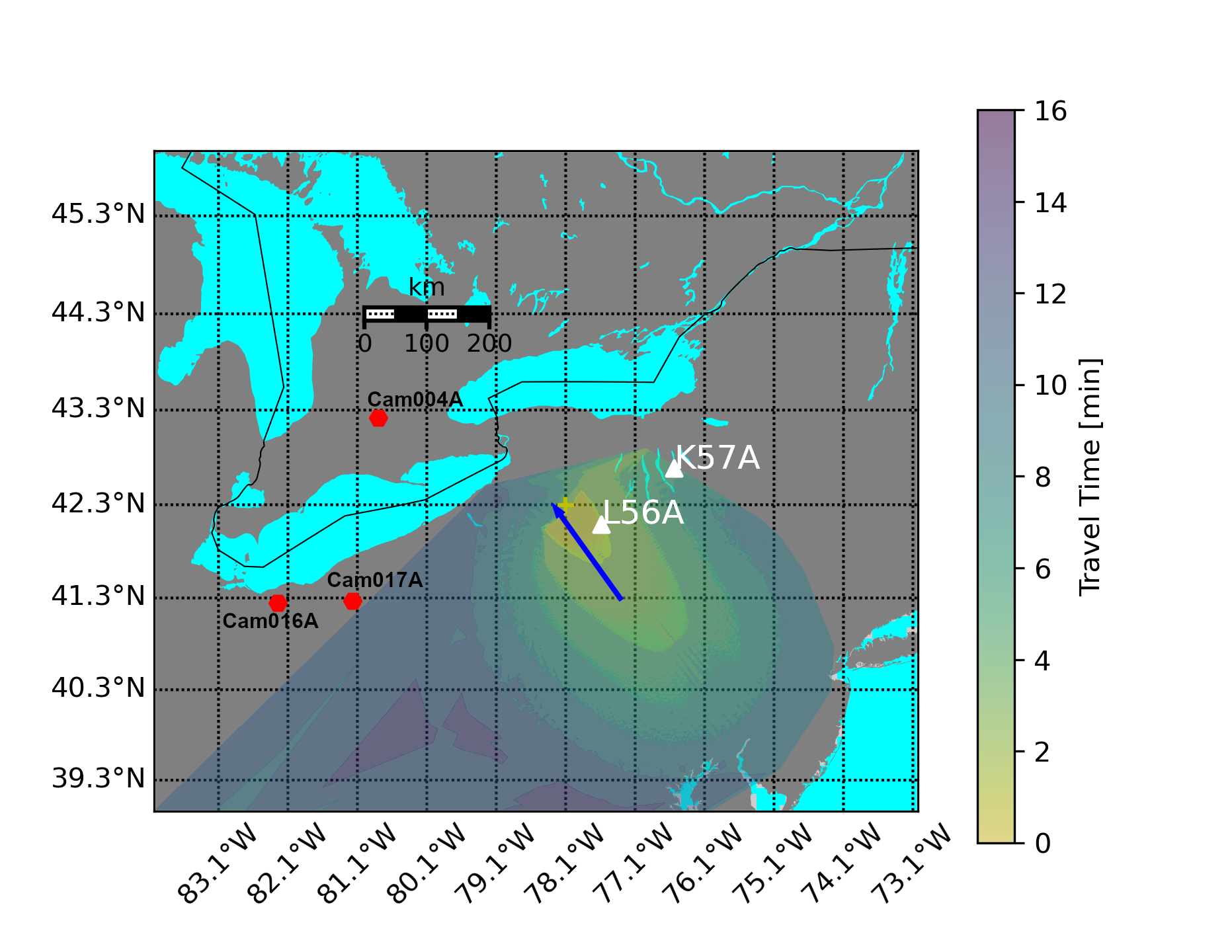}
  \caption{Trajectory of the New York state fireball (blue line) with the corresponding boom corridor (shaded region). The infrasound stations (white triangles) used in our analysis are well-positioned around the trajectory. The trajectory was found using three All-Sky cameras (red hexagons). Note that Cam16A was only used for the trajectory, and not for the photometry data. Note that the boom corridor extends far to the South West because the winds allowed more arrivals to fall within this region.}
  \label{fig:ny_map}
\end{figure*}

\begin{figure*}
  \includegraphics[width=\linewidth]{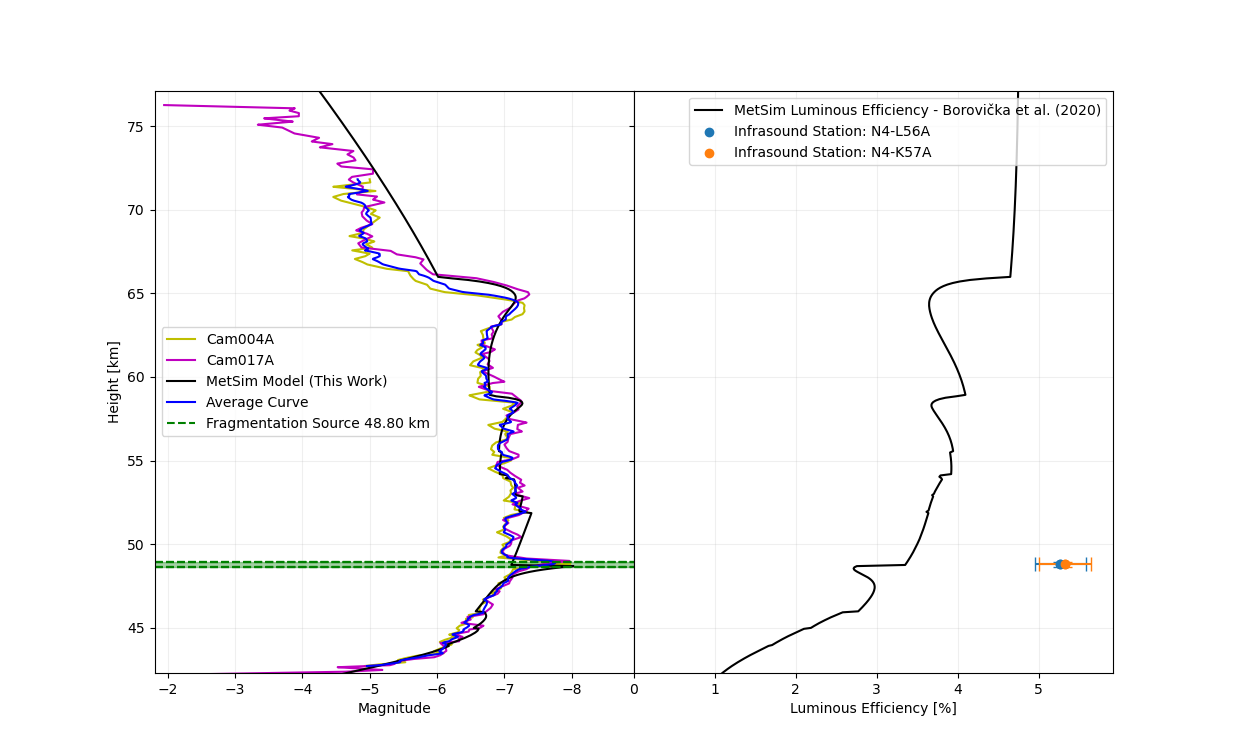}
  \caption{The measured absolute magnitude of the New York state fireball as a function of height (left plot). Cameras 016A and 017A had contamination from local light behind the meteor, which made it difficult to isolate the light emitted by the meteoroid from the background noise. Photometry was done following the procedure summarized in \citet{Vida2021}. The right plot shows the entry model mass-loss weighted luminous efficiency (dark line) while the equivalent luminous efficiency from each camera for each of the two infrasound stations are shown by the colored triangles. Color code per camera is shown in the left hand plot.}
  \label{fig:ny_lc_wmpl}
\end{figure*}

\begin{figure*}
  \includegraphics[width=\linewidth]{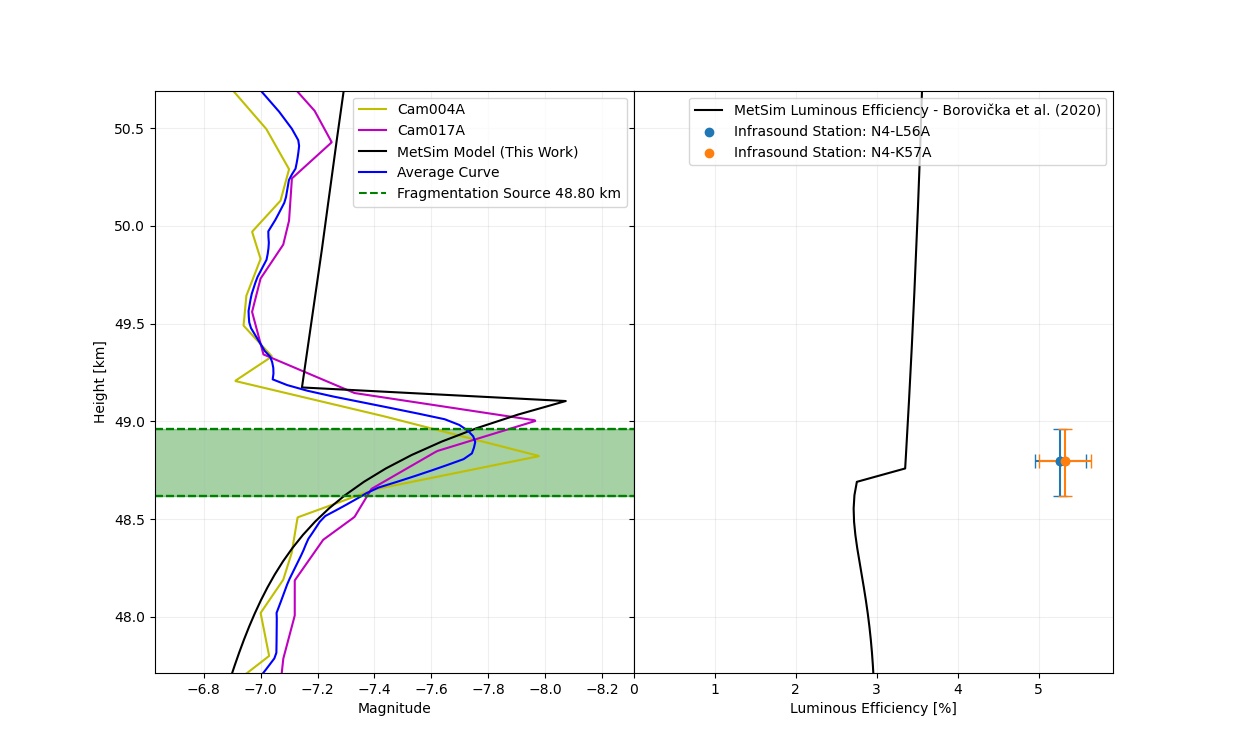}
  \caption{An enlarged version of Figure \ref{fig:ny_lc_wmpl} centred around the time of the fragmentation.}
  \label{fig:ny_lc_wmpl_zoom}
\end{figure*}



We used the length of the acoustic waveform to determine the approximate length of the fragmentation. The begin and end timings of the waveform were ray-traced to find their corresponding heights. This gave the length of the fragmentation which was moved to the flare on the light curve. \par

\begin{figure}
  \includegraphics[width=\linewidth]{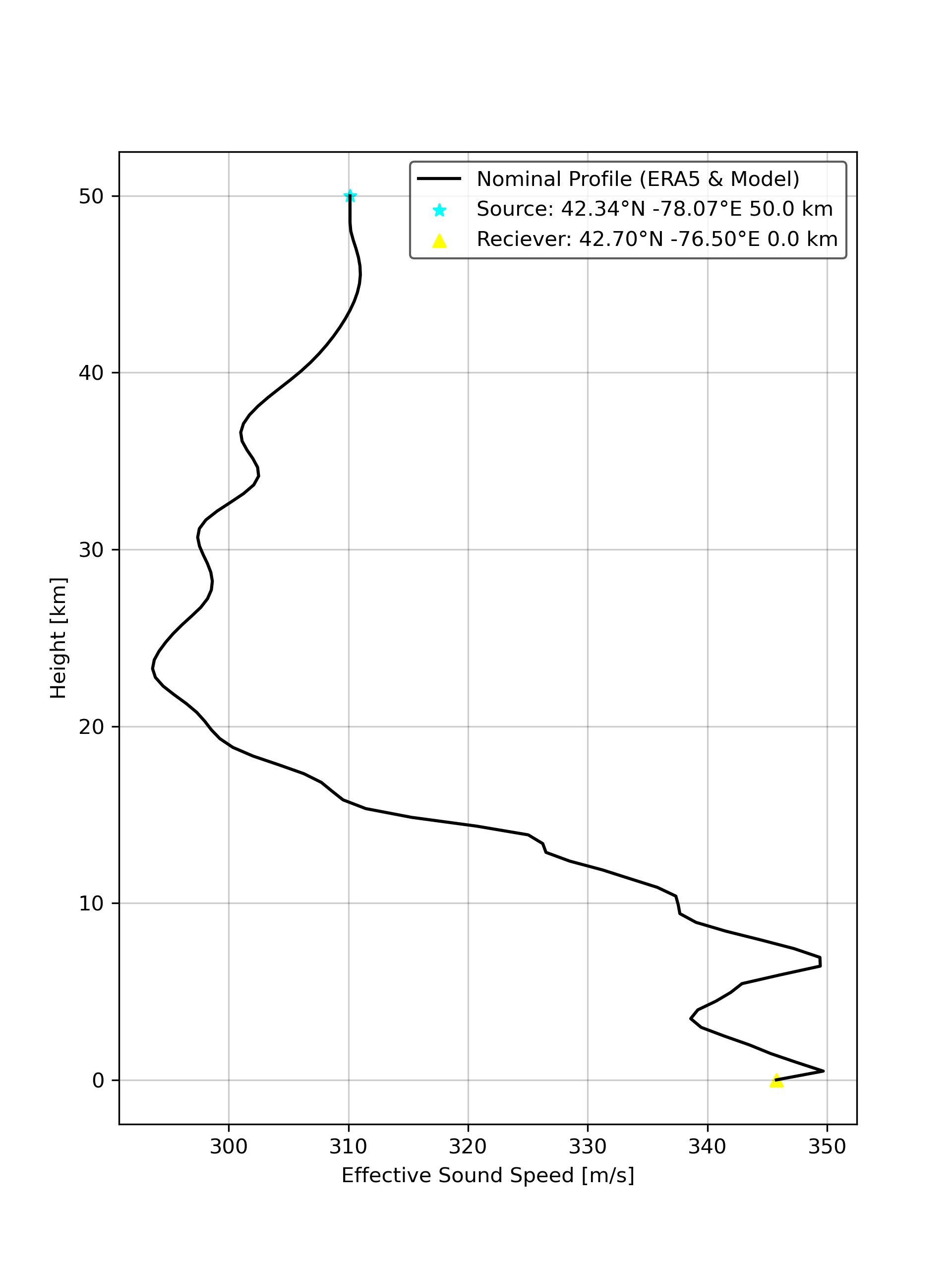}
  \caption{The effective sound speed (temperature and wind effects on the speed of sound) of a ray travelling from the trajectory to the N4-K57A station. We note the strong increase in speed for rays below~20 km, which indicates that a ray launched at speeds of $\approx 300$~m/s would likely be reflected upwards and not present at this station.}
  \label{fig:ny_eff_sound_speed}
\end{figure}

The atmospheric effective sound speed profile did not allow for lower-source acoustic rays to reach stations, as the sharp increase in sound speed at around 20~km in height, as shown in Figure \ref{fig:ny_eff_sound_speed}, would cause rays to reflect upwards. This would affect rays originating from a slower sound speed than the peak observed at around 20~km in height.

Computed luminous efficiencies using optical energies computed from individual cameras are listed in Table \ref{NY_lum_eff_data}.  The two camera stations with reliable photometry (004A and 017A) produce luminous efficiencies very similar to those predicted by the \citet{Borovicka2020} $\tau$-model. 


\section{Discussion}
\subsection{Summary of Findings for each Fireball} \label{discussion_summary}
\begin{table*}[]
\begin{tabular}{@{}llllllll@{}}
\toprule
Event                    & \begin{tabular}[c]{@{}l@{}}Height \\ {[}km{]}\end{tabular} & Source & \begin{tabular}[c]{@{}l@{}}$\tau_{M}$ {[}\%{]} \\ (Model)\end{tabular}  & \begin{tabular}[c]{@{}l@{}}$\tau_{Ac}$ {[}\%{]}\\ (This Work)\end{tabular} & $ \tau_{Ac} - \tau_{M} [\%]$ & \begin{tabular}[c]{@{}l@{}} Acoustically Derived \\ Energy {[}kT TNT{]} \end{tabular} & \begin{tabular}[c]{@{}l@{}} CNEOS Derived \\ Energy {[}kT TNT{]}\end{tabular} \\ \midrule
\multirow{3}{*}{Romania}
                         & $41.9^{+1.5}_{-0.4}$ & F & 5.63 & 5.60& 0.03 & 0.360 & \multirow{3}{*}{0.40}\\
                         & $47.7^{+0.1}_{-2.4}$ & F      & 6.95 & 9.40 & 2.45 & 0.074 &\\
                         & $55.4^{+0.9}_{-2.7}$ & F      & 9.41 & 8.40 & 1.01 & 0.013 & \\ \midrule
Alaska                   & $33.7^{+1.4}_{-5.6}$ & F & 3.58 & $6.2 \pm 0.5$ & 2.6 & $0.51^{+0.05}_{-0.05}$ & 0.53 \\ \midrule
\multirow{3}{*}{Fröslunda}  & $43.8 \pm 1.4$  & B      & 17.5  & 15.6-18.6 & -1.9 - 1.1 & $9.11 \times 10^{-5}$ /m & \multirow{3}{*}{0.33}\\
                         & $49.9^{+0.0}_{-1.4}$ & F      & 17.5 & 11.0 & -6.5 & $4.78 \times 10^{-3}$ & \\
                         & $51.7^{+1.5}_{-0.9}$  & F      & 17.5 & 19.8 & 2.3 & $4.42 \times 10^{-3}$ & \\ \midrule
New York                 & $48.8 \pm 0.2$ & F & 3.36 &  4.95 - 5.64 & 1.59 - 2.28 & $4.59 - 4.64 \times 10^{-7}$ & -                                                                    
\end{tabular}
\caption{A summary of the four fireballs used as case studies, comparing the parameters found acoustically with the model and independently found values. Here we compare the differential luminous efficiency values from this work and the model \citep{Borovicka2020}, and the acoustically derived energy to that of CNEOS (which assume a constant integral luminous efficiency of 8\%). The uncertainties given in the height column represent the corresponding time boundaries of the signal on the waveform. For events with multiple stations this represents the total height range among all observing stations. The uncertainties for the Alaska $\tau_{AC}$ and Acoustically Derived Energy represent the 95\% confidence interval. $\tau_{AC}$ is the acoustically measured luminous efficiency, $\tau_{M}$ is the model luminous efficiency, found by MetSim. Sources are marked either as F for Fragmentation, or B for Ballistic (trajectory-source)}
\label{summarytable}
\end{table*}

Table \ref{summarytable} provides a summary of the model-derived and acoustically determined luminous efficiencies for the four fireballs considered in this study. The uncertainty in height corresponds to the fragmentation height bounds found based on the acoustic signal duration. 


For fragmentation episodes of the Romanian fireball, the model-predicted and acoustically derived luminous efficiencies are in correspondence, typically within a factor of two. Uncertainties in these values are primarily influenced by two key factors: the duration of the fragmentation and the accuracy of the overpressures measured by the infrasound array. We assumed that the fragmentation duration was constrained by the acoustic waveform length, except for the 47.7~km fragmentation which was determined manually from the light curve. Changing our fragmentation boundaries would lead to different calculated luminous efficiencies. In relation to the infrasound array, the use of only 4 out of the 6 array elements could result in inaccurate measurements of overpressures, which was notably observed in the situation involving stations RO-IPH4 and RO-IPH7. We observed that the recorded amplitude of the stations did not correspond to the order of magnitude of the remaining 4 stations suggesting an issue with the amplitude calibration of these microphones.

We also considered a 38.4~km ballistic return, not presented in our Results. A complication of this return is that the descending portion of the lightcurve was heavily contaminated by the bright trail left by the fireball \citep{Borovicka2017} which is not modelled by the ablation model. Our acoustic energy deposition per unit path length is significantly lower than predicted by the model light curve, suggesting a lower luminous efficiency. The energy here was calculated through the \texttt{Geminus} program, assuming a full weak-shock propagation from source to ground, and a transfer to the linear regime (acoustic waves) \citep{geminus}. We find a result in the range of 1.77 - 2.09\%; very different from the 7.00\% model tau estimate. This suggests our ballistic estimate for $\tau$ is probably a lower limit in this case. However, due to the contamination of the light we are unable to verify the completeness of this result, and have opted to mention it here instead of as a more firm conclusion with our fragmentation results.


For the Alaska event, the results are heavily dependent on the accuracy of the height and timing information provided by USG sensors. Using the provided height of 33.7~km for the main fragmentation, an energy of $0.51^{+0.05}_{-0.04}$ kT TNT was calculated, corresponding to an integral luminous efficiency of 6.2\%. This result matches well with the data reported by CNEOS, which had a fixed integral luminous efficiency of 8\%. Allowing for uncertainty in the height of peak brightness, acoustic signal arrivals produced a supracenter solution with an optimal height of 30.3~km. This resulted in yields over a range of $0.46^{+0.40}_{-0.24}$ kT TNT and luminous efficiencies ranging from 3.67\% to 14.3\%. Such large uncertainties are to be expected for a two-station-only solution. Examining the luminous efficiency around the fragmentation, a value of 6.20\% is found. The large release of dust during the fragmentation results in the model mass-weighted differential luminous efficiency changing rapidly from 7\% to 3\% over just a few kilometers of height (3.58\% nominal at the reported fragmentation height). Within this wide bound, our results match well to the luminous efficiency model proposed by \cite{Borovicka2020}.


The Fröslunda event is the first instrumentally recorded iron meteorite-producing fireball. As such, modelling it with the \citet{Borovicka2020} luminous efficiency model which is calibrated on chondritic fireballs is not appropriate. CNEOS reports a yield of 0.33~kT TNT, assuming an 8\% integral luminous efficiency and 6000~K black body. Far-field acoustic measurements all suggest a much smaller yield by orders of magnitude, indicating either a very high integral luminous efficiency or a breakdown in the 6000~K black body assumption used to generate the original optical energy estimate \citep[][]{Tagliaferri94}. Similarly, the luminous efficiencies found for the acoustic arrivals identified as distinct fragmentations were all with a factor of two of our limited constant model. Determining the height boundaries of each flare from the light curve was challenging; we assumed that pressure waves observed at the station originated from the feature on the light curve with a height range delineated by the signal duration in infrasound station timing. This range would be affected by errors in the length of the waveform, as well as acoustic ray patterns not modelled in the ray-tracing, such as ray spreading and scattering. \par

As modelling the fireball with the \citet{Borovicka2020} model is not appropriate, we opted for a simple constant $\tau$ model. We initially modelled the light curve to close agreement with the observed lightcurve adopting an initial estimate of $\tau = 12\%$. The final mass was on the order of 100~kg, much larger than the recovered mass of 13.8~kg \citep{Kyrylenko2023}. After modelling the acoustic luminous efficiencies, we found that a better estimate was closer to 17.5\%. This resulted in a final model mass of 47.4~kg. Further minor adjustments can be made to this model, such as to the luminous efficiency and the model fragmentation parameters to obtain a better light curve fit while keeping the final mass constrained on the order of 13.8~kg. We fix tau to 17.5\% for simplicity and consistency with our acoustic values. We expect the tau modelling will change once the constant tau assumption is removed with more precise future observations of iron fireballs.

Theoretical model estimates for the luminous efficiency of iron fireballs \citep{Golub1997} suggest some difference is expected compared to chondritic compositions, of about an order of magnitude, but this model suggests differences between irons and chondritic impactors for the height and speed for Fröslunda that are only modest (of order a few percent) and much smaller than we find. We suspect that the spectral energy distribution for iron meteoroids is different compared to chondritic fireballs. Notably, the order of magnitude difference between the USG energy (which assumes a 6000~K black body) and that from far-field acoustic stations is unusual as demonstrated by \cite{Gi2017} who found only one fireball (out of 57) where multi-station periods produced such a large yield difference. 

A similar energy difference was also noted for the Apr 23, 2001 fireball in the Pacific \citep{Brown2002d}, the only other fireball to our knowledge showing such a large acoustic versus optical yield variance. For future work, we suggest computing a black body equivalent temperature for iron objects, which would help constrain the energies. Given the evidence from Fröslunda, it is possible that the differences between optical and acoustic yields indicate iron composition and might even be used to identify iron fireballs. In this regard, analysis of more infrasound records from fireballs of known iron composition, even at smaller energies, would be of great interest. 


For the New York State fireball, only two stations, N4-L56A and N4-K57A, produced consistent fragmentation energies for the flare at 48.8~km. Despite the fact that these two stations observed overpressures varying over a factor of two and at $3\times$ larger distances, almost identical yields were inverted from both. This consistency offered a test for the blast scaling laws proposed in \citep{Mcfadden2021}, as the two stations were likely to have experienced similar atmospheric profiles from the source to the station, given their alignment with the trajectory. The calculated luminous efficiency for the New York fireball was found to be less than a factor of two larger than the model. Note that we have removed the lightcurve for Cam016A from the analysis due to background light contamination which resulted in the flare appearing significantly brighter than at the other two stations. We only used this camera's data for the trajectory.

Another independent check on the mass/energy of the New York state fireball uses the approach recently proposed by \cite{Johnston2024}. In this technique, the brightness of the fireball at 60~km altitude, together with the speed, is used to estimate the initial diameter of the meteoroid. The approach presumes that there is little or no fragmentation at high altitudes and is motivated by simulations which showed that the brightness of a fireball at 60~km height is primarily determined by its size, not ablation or material characteristics. This approach requires that fragmentation is not apparent prior to 65~km altitude. This condition is just barely met for the New York state fireball and following the procedure of \cite{Johnston2024} we extrapolate the pre-fragmentation lightcurve just above 65~km to 60~km. This produces an estimated magnitude of -7 at 60~km. Using the relation in \cite{Johnston2024} we derive an estimated diameter of 0.084~m for the New York state fireball. This is slightly below the 10~cm lower diameter limit quoted by \cite{Johnston2024}, hence this result is an extrapolation outside the stated range of applicability. 

With this limitation in mind, we need to estimate equivalent mass, which requires an estimate of bulk density. For this constraint, we appeal to the PE and Pf criteria, which are one-dimensional measures of the relative strength/response to ablation \citep{Borovicka2022physical} using fireball end height and peak dynamic pressure, respectively. From the trajectory solution for the New York State fireball (see Table \ref{NY_fireball_traj_sol}) we derive a PE of -4.25 and a Pf of 1.5. Both of these values indicate a strong/chondritic-type body. Presuming the most likely association is with ordinary chondrites, we use a mean bulk density of 2100~kg m$^{-3}$ as found from Metsim modelling, suggesting significant porosity in the original meteoroid. Under this assumption, we arrive at an estimated mass of 0.65~kg using the method of \citet{Johnston2024}, very similar to our modelled value of 0.84~kg (Table \ref{metsim_ny}). 

In addition to the observations at N4-L56A and N4-K57A, there were arrivals from higher up in the trajectory at N4-M57A and IU-SSPA (around 60-70 km). These arrivals are consistent with the flare observed in the light curve at this time. The launch angle from the trajectory was such that these could be either from the ballistic source or from the fragmentation. However, we were unable to get a consistent height (within about 10~km) and energy for these stations (differing by orders of magnitude). This suggests that not all of the arrivals from these heights are direct arrivals (for example, the same time of arrival at a station could represent a direct arrival from a higher height or an indirect arrival from a lower height). The resulting uncertainty in range to the source and whether the arrival was direct or not is a likely reason for the order of magnitude difference in energies.

\subsection{Applications}

The procedure outlined in the study can be applied to any fireball where both acoustic and optical data are available. While we were only able to identify four fireballs which had sufficient acoustic and optical information, the integration of data from sources such as the dense N4 infrasound array and automated GLM detection of bolides offers the prospect for the synthesis of acoustic and optical information for many events in future and better validate existing luminous efficiency models.

Similarly, systems like the Raspberry Pi Shake \& Boom\footnote{https://raspberryshake.org/} with dense arrays in regions like the UK, along with the many independent fireball cameras operated in the region \citep{McMullan2023}, could enable similar measurements of luminous efficiency. 



\subsection{Sources of Uncertainty and Future Improvements}

The study emphasized that the energy estimation of a fragmentation is highly sensitive to both the height of the fragmentation and the overpressure measured at the station, which is related to the source energy through the blast scaling laws. To accurately estimate overpressures, bandpass filtering techniques are used to maximize the signal-to-noise ratio. Such bandpass choices naturally introduce uncertainty in the final amplitudes, though the error is generally small, provided the bandpass covers the dominant period \citep{Ens2012}. 

In the context of analyzing meteor fireballs, determining the optical energy from a fragmentation event involves calculating the area underneath the intensity versus height curve associated with the event. This area represents the total optical energy emitted by the meteor during that specific fragmentation event. However, accurately estimating the optical energy from a fragmentation can be challenging due to uncertainties in the range of heights that are involved in the fragmentation. To address this challenge, various approaches can be employed:

\begin{itemize}
    \item \textbf{Matching Luminous Efficiency to Model}: One approach, involves adjusting the luminous efficiency value to match the model. By finding the luminous efficiency value that corresponds to the model's prediction and then determining the area necessary to produce this amount of optical energy, we can estimate the bounds of the fragmentation event, and the corresponding luminous efficiency.
    \item \textbf{Percentage of Total Maximum Energy}: In cases where it is challenging to determine the heights associated with a fragmentation event precisely, we consider a certain percentage of the total maximum energy of the light curve. This approach assumes that the fragmentation contributes a certain portion of the total luminous energy, which can then be used to estimate the area under the light curve that corresponds to the fragmentation.
    \item \textbf{Acoustic Arrival Timings}: Ideally, the timing of the acoustic arrival could be used to constrain the heights of fragmentations. However, in practice, noise and overlapping signals can make it difficult to accurately determine the lengths of these phases, leading to potential underestimations of the energy.
\end{itemize}
The choice of method will depend on the specifics for each event and the quality of the data available. \par


\section{Conclusion}

In this validation study, the luminous efficiency model of \cite{Borovicka2020} was tested against four fireballs. The agreement between the calculated infrasonic energy estimates and the energies reported by CNEOS and optical data indicates that the \citet{Borovicka2020} model generally provides consistent results for both differential and integral luminous efficiencies.

In addition, we present the methodology for infrasonically determining the yield of fragmentations as an extension of the work presented in \cite{Mcfadden2021}. Using the Alaska event, we show that it may be possible to determine reasonable bounds on the yield of fragmentation without knowing its exact location and height.

In two cases, the luminous efficiency measurements showed agreement to be better than a factor of two to model values. The New York state fireball had a flare where the acoustic luminous efficiency was a factor of two higher than predicted by the \cite{Borovicka2020} model. However, the model bulk density for the fireball was also unusually low for a nominally chondritic body (2100~kg/m$^3$).

However, the most significant deviation was unsurprisingly observed for the Fröslunda fireball, produced by an iron meteoroid. The differential and integral luminous efficiency was found to be more than a factor of three higher than the \cite{Borovicka2020} model calibrated to chondritic fireballs. This suggests that iron fireball luminous efficiencies are higher than those produced by chondritic meteoroids and what past radiation models have estimated \cite{ReVelle2001}. If this finding holds true for more calibrated events, the large difference between USG lightcurve energy and infrasonically estimated source energy offers a means to potentially distinguish iron fireballs from the larger background population of chondritic impactors.

\section{Acknowledgements}
This work was supported in part by the NASA Meteoroid Environment Office under cooperative agreement 80NSSC21M0073. This research has made use of the neo-bolide.ndc.nasa.gov website, which was developed and operated by NASA's Asteroid Threat Assessment Project. .

\bibliography{main_rev2}

\appendix

\section{\texttt{MetSim} Models} \label{MetSim_models}

The data presented in the following tables are the parameters of the models generated by \texttt{MetSim} for this study. Table \ref{MetSim_init_params} gives the initial parameters, where Tables \ref{metsim_romania} through \ref{metsim_ny} give the fragmentation models for each fireball. \par
For Tables \ref{metsim_romania} to \ref{metsim_ny}, fragmentations are given as either F - fragmentation into additional bodies which undergo single body ablation, EF - fragmentation into meteoroids which ablate through erosion of grains as well as single body ablation, D - dust release, or A - parameter change of all fragments.  For each mass release, it is assumed that the number of fragments is 1, and that gamma is 1.0. \par
Fragmentation models were fit such that the model light curve matched the observed light curve, but were constrained by independent sources whenever possible.

\begin{table*}[]
\begin{tabular}{llllllllll}
\toprule
Event    & \begin{tabular}[c]{@{}l@{}}$P_0$ \\ {[}W{]}\end{tabular} & \begin{tabular}[c]{@{}l@{}}$\rho_{bulk}$ \\ {[}kg/m\textasciicircum{}3{]}\end{tabular} & \begin{tabular}[c]{@{}l@{}}$\rho_{grain}$ \\ {[}kg/m\textasciicircum{}3{]}\end{tabular} & $m_{init}$ {[}kg{]} & $\sigma$ {[}kg/MJ{]} & $v_{init}$ {[}km/s{]} & $A$  & $\Gamma$ & $Z_c$ {[}deg{]} \\
\midrule
Romania & 1500 & 2500 & 3500 & 4500 & 0.010 & 27.878 & 0.70 & 1.00 & 47.377 \\
Alaska   & 3030 & 3300  & 500 & 25000 & 0.0050  & 21.1 & 1.00 & 0.8 & 16.502 \\
Fröslunda & 3030 & 7000 & 7000 & 3540 &  0.1000 & 17.6 & 1.00 & 0.8 & 17.900\\
New York & 1200 & 2100  & 3500 & 0.85 & 0.0010 & 20.08 & 1.00  & 0.8  & 72.787 \\  
\bottomrule
\end{tabular}
\caption{The initial parameters for each event used in \texttt{MetSim} for this study. For the Romania event, this data was extracted from \cite{Borovicka2017} as much as possible. Initial mass and velocity were taken from independent sources such as CNEOS and the Norsk Meteor Nettverk when available. The parameters for the New York State fireball were extracted from the All-Sky cameras and the resulting trajectory solution (available in \ref{NY_fireball_traj_sol}).}
\label{MetSim_init_params}
\end{table*}

\begin{table*}[]
\begin{tabular}{llllllll}
\toprule
Type &
  \begin{tabular}[c]{@{}l@{}}Height \\ {[}km{]}\end{tabular} &
  \begin{tabular}[c]{@{}l@{}}Mass \\ Loss\\ {[}\%{]}\end{tabular} &
  \begin{tabular}[c]{@{}l@{}}Parent\\ Mass\\ {[}kg{]}\end{tabular} &
  \begin{tabular}[c]{@{}l@{}}Ablation\\ Coefficient\\ {[}s\textasciicircum{}2/km\textasciicircum{}2{]}\end{tabular} &
  \begin{tabular}[c]{@{}l@{}}Erosion\\ Coefficient\\ {[}s\textasciicircum{}2/km\textasciicircum{}2{]}\end{tabular} &
  \begin{tabular}[c]{@{}l@{}}Grain\\ Minimum\\ Mass {[}kg{]}\end{tabular} &
  \begin{tabular}[c]{@{}l@{}}Grain\\ Maximum\\ Mass {[}kg{]}\end{tabular} \\
  \midrule
\begin{tabular}[c]{@{}l@{}}Initial\\ Conditions\end{tabular} & 180.0  & -  & -    & 0.01  & 0.00 & -    & -     \\
EF                                                           & 79.0   & 7  & 4500 & 0.024 & 2.00 & 1e-6 & 1e-5 \\
EF                                                           & 64.0   & 2  & 4170 & 0.01  & 0.50 & 1e-5 & 1e-4 \\
EF                                                           & 55.0   & 1  & 4060 & 0.04  & 2.00 & 1e-5 & 1e-5 \\
EF                                                           & 48.5   & 25 & 3970 & 0.05  & 0.20 & 1e-3 & 1e-3  \\
EF                                                           & 44.8   & 33 & 2930 & 0.06  & 0.60 & 1e-3 & 1e-3 \\
EF                                                           & 43.3   & 99 & 1950 & 0.065 & 1.00 & 1e-3 & 1e-3  \\
End                                                          & 27.957 & - & -     & -    & -    & -    & -     \\
\bottomrule
\end{tabular}
\caption{The fragmentation model used in \texttt{MetSim} for the Romania Fireball. The parameters for the fragmentation were assumed from \cite{Borovicka2017}, with some changes in the heights and mass losses of the fireball. Here the mass distribution index was assumed to be 2.00.}
\label{metsim_romania}
\end{table*}

\begin{table*}[]
\begin{tabular}{lllllllll}
\toprule
Type &
  \begin{tabular}[c]{@{}l@{}}Height\\ {[}km{]}\end{tabular} &
  \begin{tabular}[c]{@{}l@{}}Mass\\ Loss \\ {[}\%{]}\end{tabular} &
  \begin{tabular}[c]{@{}l@{}}Parent\\ Mass \\ {[}kg{]}\end{tabular} &
  \begin{tabular}[c]{@{}l@{}}Ablation\\ Coefficient\\ {[}s\textasciicircum{}2/km\textasciicircum{}2{]}\end{tabular} &
  \begin{tabular}[c]{@{}l@{}}Erosion\\ Coefficient\\ {[}s\textasciicircum{}2/km\textasciicircum{}2{]}\end{tabular} &
  \begin{tabular}[c]{@{}l@{}}Grain\\ Minimum\\ Mass {[}kg{]}\end{tabular} &
  \begin{tabular}[c]{@{}l@{}}Grain\\ Maximum\\ Mass {[}kg{]}\end{tabular} &
  \begin{tabular}[c]{@{}l@{}}Mass\\ Index\end{tabular} \\
  \midrule
\begin{tabular}[c]{@{}l@{}}Initial\\ Conditions\end{tabular} & 180.0  & -  & -    & 0.005 & 0.00 & -    & -    & -    \\
D                                                            & 35.0   & 50  & 24600 & 0.005 & 0.00 & 1e-2 & 5 & 2.5 \\
EF$\times6$                                                           & 34.5   & 95 & 12300 & 0.005 & 0.73 & 1e-3 & 5 & 2.5 \\
End                                                          & 12.010 & -  & 244  & -     &  -    & -    & -    & - \\
\bottomrule
\end{tabular}
\caption{The fragmentation model used in \texttt{MetSim} for the Alaska Fireball.}
\end{table*}

\begin{table*}[]
\begin{tabular}{llllllll}
\toprule
Type &
  \begin{tabular}[c]{@{}l@{}}Height \\ {[}km{]}\end{tabular} &
  \begin{tabular}[c]{@{}l@{}}Mass \\ Loss\\ {[}\%{]}\end{tabular} &
  \begin{tabular}[c]{@{}l@{}}Parent\\ Mass\\ {[}kg{]}\end{tabular} &
  \begin{tabular}[c]{@{}l@{}}Ablation\\ Coefficient\\ {[}s\textasciicircum{}2/km\textasciicircum{}2{]}\end{tabular} &
  \begin{tabular}[c]{@{}l@{}}Erosion\\ Coefficient\\ {[}s\textasciicircum{}2/km\textasciicircum{}2{]}\end{tabular} &
  \begin{tabular}[c]{@{}l@{}}Grain\\ Minimum\\ Mass {[}kg{]}\end{tabular} &
  \begin{tabular}[c]{@{}l@{}}Grain\\ Maximum\\ Mass {[}kg{]}\end{tabular} \\
  \midrule
\begin{tabular}[c]{@{}l@{}}Initial\\ Conditions\end{tabular} & 180.0  & -  & -    & 0.100  & 0.00 & -    & -     \\
EF & 66.0  & 16  & 3530    & 0.200  & 0.90 &1e-4   & 1e-3   \\
A & 43.5  & -  & 2760  & 0.040  & -&  -   &  -    \\
EF & 40.0  & 
8 & 2710   & 0.100  & 0.05 & -   & -    \\
A & 24.0  & -  & 1490   & 0.038  & - & -    & -    \\
A & 23.5  & -  & 1430   & 0.036  & -& -    & -     \\
A & 23.0  & -  & 1370    & 0.034  & - & -    & -     \\
A & 22.5  & -  & 1300   & 0.032  & - & -    & -     \\
A & 22.0  & -  & 1220    & 0.030  & - & -    & -    \\
A & 21.0 & -  & 1100   & 0.028 & - & -    & -     \\
A & 20.0  & -  & 956    & 0.026  & - & -   & -     \\
A & 19.0  & -  & 841    & 0.024  & - & -    & -     \\
End & 7.953  & -  & 47.4   & -  & - & -    & -     \\
\bottomrule
\end{tabular}
\caption{The fragmentation model used in \texttt{MetSim} for the Fröslunda Fireball.}
\end{table*}

\begin{table*}[]
\begin{tabular}{llllllll}
\toprule
Type &
  \begin{tabular}[c]{@{}l@{}}Height\\ {[}km{]}\end{tabular} &
  \begin{tabular}[c]{@{}l@{}}Mass\\ Loss \\ {[}\%{]}\end{tabular} &
  \begin{tabular}[c]{@{}l@{}}Parent\\ Mass \\ {[}kg{]}\end{tabular} &
  \begin{tabular}[c]{@{}l@{}}Ablation\\ Coefficient\\ {[}s\textasciicircum{}2/km\textasciicircum{}2{]}\end{tabular} &
  \begin{tabular}[c]{@{}l@{}}Erosion\\ Coefficient\\ {[}s\textasciicircum{}2/km\textasciicircum{}2{]}\end{tabular} &
  \begin{tabular}[c]{@{}l@{}}Grain\\ Minimum\\ Mass {[}kg{]}\end{tabular} &
  \begin{tabular}[c]{@{}l@{}}Grain\\ Maximum\\ Mass {[}kg{]}\end{tabular} \\
  \midrule

\begin{tabular}[c]{@{}l@{}}Initial\\ Conditions\end{tabular} & 180.0  & -  & -    & 0.001  & 0.00 & -    & -     \\
EF & 66.0  & 12  & 0.841    & 0.005  & 0.00 & 1e-5    & 1e-3     \\
EF & 59.0  & 5  & 0.727    & 0.005  & 0.5 & 1e-5    & 1e-3     \\
D & 55.6  & 0.4  & 0.680    & 0.005  & 0.4 & 1e-5   & 1e-3     \\
EF & 54.2  & 5  & 0.672    & 0.005  & 0.17 & 1e-4    & 1e-3     \\
A & 54.0  & -  & 0.638    & 0.002  & 0.03 & -    & -     \\
A & 53.0  & -  & 0.630    & 0.003  & 0.03 & -    & -     \\
A & 52.0  & -  & 0.618    & 0.005  & 0.03 & -    & -     \\
D & 49.2  & 12  & 0.562    & 0.005  & 0.03 & 1e-5    & 1e-4     \\
EF & 46.0  & 30  & 0.428    & 0.005  & 0.03 & 1e-4    & 1e-3     \\
EF & 45.0  & 30  & 0.284    & 0.005 & 0.03 & 1e-4    & 1e-3     \\
EF & 44.0  & 30  & 0.186    & 0.005  & 0.03 & 1e-4    & 1e-3     \\
End & 6.350  & -  & 0.0943    & -  & - & -    & -     \\
\bottomrule
\end{tabular}
\caption{The fragmentation model used in \texttt{MetSim} for the New York State Fireball. Here the mass distribution index was assumed to be 2.00.}
\label{metsim_ny}
\end{table*}

\section{Seismic and Infrasound Station References}

The seismic and infrasonic stations used in this study are given in Table \ref{station_table}.
\begin{table*}[]
\begin{adjustbox}{width=\textwidth}
\begin{tabular}{@{}llllll@{}}
\toprule
Event &
  \begin{tabular}[c]{l} Network \\ Code \end{tabular} &
  \begin{tabular}[c]{l} Stations \\ Used \end{tabular} &
  \begin{tabular}[c]{l} Latitude \\ {[}deg N{]} \end{tabular} &
  \begin{tabular}[c]{l} Longitude \\ {[}deg E{]} \end{tabular}&
  Citation \\ \midrule
\begin{tabular}[c]{@{}l@{}}Romania\\ Fireball\end{tabular} &
  RO &
  \textbf{IPH Array} &
  45.9 &
  26.6 &
  \cite{STAT-RO} \\ \midrule
\multirow{2}{*}{\begin{tabular}[c]{@{}l@{}}Alaska \\ Fireball\end{tabular}} &
  AK &
  COLD &
  67.2 &
  -150.2 &
  \cite{STAT-AK} \\
 &
  TA &
  \textbf{TOLK} &
  68.6 &
  -149.6 &
  \cite{STAT-TA} \\ \midrule
\multirow{2}{*}{\begin{tabular}[c]{@{}l@{}}Fröslunda \\ Fireball\end{tabular}} &
  AM &
  \begin{tabular}[c]{@{}l@{}}R3993*\\ \textbf{R796B*}\\ RCC91*\\ R65C9\\ RACBA\end{tabular} &
  \begin{tabular}[c]{@{}l@{}}59.5\\ 59.5\\ 59.5\\ 59.2\\ 59.3\end{tabular} &
  \begin{tabular}[c]{@{}l@{}}18.3\\ 18.3\\ 18.3\\ 18.4\\ 18.1\end{tabular} &
  \cite{STAT-AM} \\
 &
  \multicolumn{2}{l}{\textbf{\begin{tabular}[c]{@{}l@{}}I26DE-Freyung, Germany\\ I37NO-Bardufoss, Norway\end{tabular}}} &
  \begin{tabular}[c]{@{}l@{}}48.9\\ 69.1\end{tabular} &
  \begin{tabular}[c]{@{}l@{}}13.7\\ 18.6\end{tabular} &
  \cite{STAT-IS} \\ \midrule
\begin{tabular}[c]{@{}l@{}}New York\\ State \\ Fireball\end{tabular} &
  IU &
  \textbf{SSPA} &
  40.6 &
  -77.9 &
  \cite{STAT-IU} \\
 &
  N4 &
  \textbf{\begin{tabular}[c]{@{}l@{}}K57A\\ L56A\\ M57A\end{tabular}} &
  \begin{tabular}[c]{@{}l@{}}42.7\\ 42.1\\ 41.3\end{tabular} &
  \begin{tabular}[c]{@{}l@{}}-76.5\\ -77.6\\ -77.1\end{tabular} &
  \cite{STAT-N4} \\ \bottomrule
\end{tabular}
\end{adjustbox}
\caption{The seismic and infrasonic stations used for each case study. Stations in bold include an infrasound component, and stations with an asterisk are co-located. For seismic stations, channels ending in HZ where used (such as HHZ), in order to get the vertical component of motion. For infrasonic stations, either the BDF or HDF channels were used depending on what is available.}
\label{station_table}
\end{table*}

\section{New York State Fireball Solution} \label{NY_fireball_traj_sol}
Parameters of the New York State fireball can be found in Table \ref{NY_params}.

\begin{table}[]
\begin{tabular}{@{}cllll@{}}
\toprule
\multicolumn{1}{l}{} &
   &
  Value & 95\% Confidence Interval \\
  Unit \\ \midrule
\multicolumn{1}{l}{Begin Point} &
  \begin{tabular}[c]{@{}l@{}}Latitude\\ Longitude\\ Height \\ Time\end{tabular} &
  \begin{tabular}[c]{@{}l@{}}41.3198\\ -77.2898\\ 75.779 \\ 01:45:51.671040\end{tabular} &
  \begin{tabular}[c]{@{}l@{}}  (41.3186, 41.3197)\\ (-77.2886,  -77.2513) \\ (75.700,    76.348)\\ \\\end{tabular} &
  \begin{tabular}[c]{@{}l@{}}deg N\\ deg E\\ km \\ UTC\end{tabular} \\\midrule
\multicolumn{1}{l}{End Point} &
  \begin{tabular}[c]{@{}l@{}}Latitude\\ Longitude\\ Height \end{tabular} &
  \begin{tabular}[c]{@{}l@{}}42.2882\\ -78.2697\\ 35.445\end{tabular} &
  \begin{tabular}[c]{@{}l@{}} (42.2827,   42.3008)\\ (-78.3014,  -78.2563)\\   (34.805,    35.702) \end{tabular} &
  \begin{tabular}[c]{@{}l@{}}deg N\\ deg E\\ km\end{tabular}\\\midrule
  \multicolumn{1}{l}{Radiant (ECI)} &
  \begin{tabular}[c]{@{}l@{}}Right Ascension\\ Declination \\ Azimuth \\ Elevation \\ $v_{\mathrm{avg}}$ \\ $v_{\mathrm{init}}$ \end{tabular} &
  \begin{tabular}[c]{@{}l@{}}314.51\\ -22.70\\ 144.04 \\ +17.17\\ 17.82\\ 19.67\end{tabular} &
  \begin{tabular}[c]{@{}l@{}}(314.40, 315.94) \\ (-22.74, -22.07) \\(142.54, 144.17)\\ (+17.00, +17.27)\\ (17.81,  18.46)\\ (19.66,  20.08)\end{tabular}&
  \begin{tabular}[c]{@{}l@{}}deg \\ deg \\ deg\\ deg\\ km/s\\ km/s\end{tabular}  \\\midrule
  \multicolumn{1}{l}{Radiant (Geocentric)} &
  \begin{tabular}[c]{@{}l@{}}Right Ascension\\ Declination  \\ $v_{g}$ \\ $v_{\mathrm{inf}}$ \end{tabular} &
  \begin{tabular}[c]{@{}l@{}}318.65\\ -29.81\\ 16.22\\ 19.67\end{tabular} &
  \begin{tabular}[c]{@{}l@{}} (318.54, 319.97) \\ (-29.87, -28.80)\\ (16.21,  16.72)\\ (19.66,  20.08)\end{tabular} &
  \begin{tabular}[c]{@{}l@{}}deg \\ deg\\ km/s\\ km/s\end{tabular} \\\midrule
\multicolumn{1}{l}{\begin{tabular}[c]{@{}l@{}}State Vector\\ (ECI)\end{tabular}} &
  \begin{tabular}[c]{@{}l@{}}$v_x$\\ $v_y$\\ $v_z$\end{tabular} &
  \begin{tabular}[c]{@{}l@{}}12.719\\  -12.940 \\ -7.590\end{tabular} &
  \begin{tabular}[c]{@{}l@{}} (12.682,    13.373)\\  (-12.997,   -12.903)\\ (-7.606,    -7.526)\end{tabular}&
  \begin{tabular}[c]{@{}l@{}}km/s\\ km/s\\ km/s\end{tabular} \\\midrule
\multirow{6}{*}{Orbital Parameters} &
  $a$ &
  2.314  &  (2.302,   2.338) &
  AU \\
 &
  $e$ &
  0.662 &   (0.662,   0.671) &
   \\
 &
  $i$ &
  6.335& (6.229,   6.366) &
  deg \\
 &
  $\Omega$ &
  326.2394 & (326.2389, 326.2395) &
  deg \\
 &
  $\omega$ &
  64.706 & (64.532,  66.991) &
  deg \\
 &
  $\Pi$ &
  30.945 & (30.771,  33.231) &
  deg \\ \cmidrule(l){1-5} 
\end{tabular}
\label{NY_params}
\caption{Trajectory and orbit parameters of the New York State fireball. Physical characteristics of the meteoroid body can be found in Table \ref{MetSim_init_params}. The uncertainties presented here represent the 95\% confidence interval from 100 Monte Carlo simulations.}
\end{table}

\end{document}